\documentclass[a4paper,fleqn]{cas-sc}

\usepackage{amssymb}
\usepackage{amsmath}

\usepackage{adjustbox}
\usepackage{color}
\usepackage{xcolor}
\usepackage{bm}
\usepackage{cases}
\usepackage{subcaption}
\usepackage[numbers]{natbib}
\usepackage{natbib}

\def\tsc#1{\csdef{#1}{\textsc{\lowercase{#1}}\xspace}}
\tsc{WGM}
\tsc{QE}
\tsc{EP}
\tsc{PMS}
\tsc{BEC}
\tsc{DE}

\begin{document}
\let\WriteBookmarks\relax
\def\floatpagepagefraction{1}
\def\textpagefraction{.001}
\def\mytitle{CFD-based design optimization of a $5~\text{kW}$ ducted hydrokinetic turbine with practical constraints}
\shorttitle{\mytitle}
\shortauthors{Park et~al.}

\title [mode = title]{\mytitle}

\author[inst1]{Jeongbin Park}
\author[inst2]{Marco Mangano}
\author[inst2]{Sabet Seraj}
\author[inst2]{Bernardo Pacini}
\author[inst2]{Yingqian Liao}
\author[inst3]{Bradford G. Knight}
\author[inst1]{Kartik Naik}
\author[inst1]{Kevin J. Maki}
\author[inst2]{Joaquim R. R. A. Martins}
\author[inst1]{Jing Sun}
\author[inst1]{Yulin Pan\corref{cor1}}[orcid=0000-0002-7504-8645]
\ead{yulinpan@umich.edu}
\cortext[cor1]{Corresponding author}

\affiliation[inst1]{organization={Naval Architecture and Marine Engineering, University of Michigan},
            city={Ann Arbor},
            postcode={MI 48109}, 
            state={MI},
            country={USA}}
\affiliation[inst2]{organization={Aerospace Engineering, University of Michigan},
            city={Ann Arbor},
            postcode={48109}, 
            state={MI},
            country={USA}}

\affiliation[inst3]{organization={Ocean Engineering, University of Rhode Island},
            city={Narragansett},
            postcode={02882}, 
            state={RI},
            country={USA}}

\begin{abstract}
Ducted hydrokinetic turbines enhance energy-harvesting efficiency by better conditioning the flow to the blades, which may yield higher power output than conventional freestream turbines for the same reference area. 
In this work, we present a ducted hydrokinetic turbine design obtained by simultaneously optimizing the duct, blade, and hub geometries. 
Our optimization framework combines a CFD solver, an adjoint solver, and a gradient-based optimizer to efficiently explore a large design space, together with a feature-based parameterization method to handle the complex geometry. 
Practical geometrical constraints ensure the manufacturability of the duct in terms of a minimum thickness and the housing of a $5~\text{kW}$ generator within the hub.
The optimization converges to a short, thin duct with a rounded leading edge and an elongated hub protruding the duct inlet. 
The optimized ducted turbine achieves up to 50\% efficiency when evaluated by RANS/URANS solvers despite a bulky hub, outperforming the 45\% efficiency of the freestream Bahaj turbine featuring the same hub. 
This work showcases the effectiveness of CFD-based optimization in advancing ducted turbine designs and demonstrates the hydrodynamic benefits of a ducted configuration, paving the way for future research and real-world applications.
\end{abstract}


\begin{highlights}
\item We perform CFD-based optimization of a $5~\text{kW}$ ducted hydrokinetic turbine.
\item Design Variable Parametrization includes duct, hub, and blade shapes.
\item Practical constraints include duct leading-edge curvature and thickness, and hub packaging.
\item Optimized design achieves up to 50\% efficiency.
\item Optimized geometry features a thin cambered duct and elongated hub.
\end{highlights}

\begin{keywords}
ducted hydrokinetic turbine \sep CFD \sep gradient-based optimization \sep adjoint method


\end{keywords}

\maketitle

\section{Introduction}
\label{sec:Intro}

Hydrokinetic turbines have gained attention over the past few decades as a promising technology for harnessing renewable energy from water currents.
They offer a sustainable solution to the growing global energy demand.
One key advantage of hydrokinetic energy resources is their predictable and concentrated nature, enabling continuous power generation.
Furthermore, hydrokinetic turbines require minimal infrastructure and have limited impact~\cite{bahaj2003fundamentals,jacobson2012assssment,ibrahim2021hydrokinetic}.
Due to these advantages, hydrokinetic turbines have been researched extensively, covering various topics such as potential site assessments~\cite{kirby2022assessments}, wake flow dynamics~\cite{nago2022literature}, and environmental impact analyses~\cite{yuce2015hydrokinetic}.
Horizontal-axis hydrokinetic turbines have particularly received more attention than other turbine arrangements because of their technological maturity~\cite{lago2010advances}.

An extensively studied topic for horizontal-axis turbines is the improvement of their energy-harvesting efficiency.
One approach of growing interest is the use of a duct (also referred to as a shroud or diffuser) to condition the flow passing the turbine blades for improved efficiency~\cite{nunes2020systematic,khan2024performance}. 
In our previous paper~\cite{park2023cfd}, we reviewed different approaches for assessing the performance of ducted turbines, including models based on 1D momentum theory and its extensions~\cite{lawn2003optimization,van2007science,jamieson2009beating,werle2008ducted,bontempo2020potential}, experiments~\cite{scherillo2011numerical,shahsavarifard2015effect,roshan2015rans,shi2015optimal,coiro2016diffuser,song2019numerical,nunes2019experimental}, and computational fluid dynamics (CFD) models.

CFD models for ducted turbines include Reynolds-averaged Navier--Stokes (RANS) and unsteady RANS (URANS).
One option to model the rotor is to represent its effects using body-force-based methods, such as actuator disk models~\cite{gaden2010numerical,shives2012developing,shi2015optimal,fleming2016analysis} or blade element momentum theory (BEMT)~\cite{belloni2017investigation,allsop2017hydrodynamic}.
Another option with higher fidelity is to resolve the full 3D geometry of turbine blades within a computational mesh, where their rotation is modeled by multiple reference frame (MRF)~\cite{roshan2015rans,silva2018new,song2019numerical,rezek2023novel} or rotating-sliding mesh (RS)~\cite{tampier2017numerical,knight2018coupling}.
These advancements in ducted turbine modeling enable a more accurate assessment of their performance.
However, previous studies' assessments focused on single or few turbine designs, and thus, the performance of these designs is likely far from optimal.

Developing simulation-based optimizations for ducted turbines is a considerably more challenging task, especially when the optimization is coupled with higher-fidelity simulation methods.
Several RANS-based studies use gradient-free optimization methods, such as genetic algorithms and simulated annealing~\cite{coiro2016diffuser,barbaric2020investigation,rezek2021design}.
While these efforts provided some insights into the design of ducted turbines, they only included a few design variables (up to 14) and did not necessarily achieve the optimality.
These shortcomings stem from the use of gradient-free optimization methods, which converge slowly and do not check for the mathematical optimality conditions~\cite[Ch.~7]{Martins2022}.
As a result, these design efforts optimize only one component of the ducted turbine (e.g., duct only) using a small number of design variables, leaving room for further improvement. 

Our previous work~\cite{park2023cfd} presented the first gradient-based optimization of a ducted hydrokinetic turbine, with the gradients efficiently computed by an adjoint method coupled with a RANS solver and the MRF method~\cite{park2023cfd}.
This methodology enables simultaneous optimization of the blade and duct geometries (21 design variables in total), yielding an efficiency of approximately 54\% when re-evaluated using a high-fidelity URANS solver coupled with the RS approach.
This efficiency is significantly higher than that of the standard freestream Bahaj model, which yields 46\% efficiency~\cite{bahaj2007power}. Despite the novel contribution, the formulation of our previous work had several limitations.
First, the design excluded a hub, which is a necessary component for housing a generator for power transmission and attaching blades in actual turbines.
Secondly, while several studies have proposed a foil-shaped duct cross-section~\cite{shives2012developing,coiro2016diffuser,tampier2017numerical,do2018approach,song2019numerical,barbaric2020investigation,rezek2021design}, our previous duct design was restricted to a thin-walled shape due to limitations in the geometric parametrization approach.

The purpose of this work is to develop a more comprehensive design optimization of a ducted hydrokinetic turbine, overcoming previous limitations.
To this end, we parametrize the ducted turbine geometry with a feature-based parametric CAD software, Engineering Sketch Pad (ESP)~\cite{Haimes2013a}, which has been successfully used in gradient-based optimization of simpler, single-component geometries~\cite{Hajdik2023a,Hajdik2023b}. 
The ESP parametrization offers more direct and convenient control based on user-defined features than the previously used free-form deformation (FFD) approach~\cite{Sederberg1986,Kenway2010b}.
These features can include coordinates, polynomial coefficients, length, radius, and sectional twist, which are then connected by cubic spline surfaces to form a smooth geometry~\cite{dannenhoffer2024overview}.
ESP enables more precise adjustments of the global and local shape, addressing the limitations associated with the FFD method in our previous work.

We use this updated framework to perform a gradient-based optimization of a $5~\text{kW}$ ducted hydrokinetic turbine.
A total of 37 design variables control the geometry of the blades, hub, and duct.
In particular, the duct can have an arbitrary thickness profile constrained by a minimum thickness and a minimum leading edge radius of curvature.
The minimum thickness is a manufacturing and structural consideration, and the minimum leading edge radius is beneficial for off-design oblique flow conditions~\cite{Madsen2019a,Mangano2021a}. 
The hub shape is allowed to change only ahead and behind a central cylindrical section that remains fixed to accommodate a generator sized from electrical considerations.
In the optimization process, the efficiency of a design and its gradients are evaluated using DAFoam~\cite{He2020b}, which uses an adjoint method for gradient computation.
The optimization is performed by SNOPT, which is an implementation of sequential quadratic programming~\cite{Gill2005a}.

The optimized design yields an efficiency of $50\%$ (and $48\%\sim49\%$ after re-evaluation by a higher-fidelity URANS solver in OpenFOAM~\cite{jasak2007openfoam}) despite the bulky hub, which negatively impacts the turbine's hydrodynamic performance.
This efficiency compares favorably to $32\%$ of our baseline design of choice and $45\%$ of a freestream Bahaj turbine with the same hub.
The final design exhibits several distinct features that have not been identified before, including an elongated hub protruding from the duct inlet, a thin duct at the boundary of the constraints, and correspondingly changed blade geometry.

The paper is organized as follows.
We first define and formulate the optimization problems in Section~\ref{sec:Probstate}.
In Section~\ref{sec:Method}, we describe the methods used for the optimization and CFD evaluation, with a focus on the ESP parametrization and its integration into the optimization framework.
The results of the optimization and CFD re-evaluations are presented and discussed in Section~\ref{sec:Results}.
Finally, the conclusions are provided in Section~\ref{sec:conclusion}.

\section{Problem Statement}
\label{sec:Probstate}

\begin{figure}[h!]
\centerline{\includegraphics[width=0.5\columnwidth]{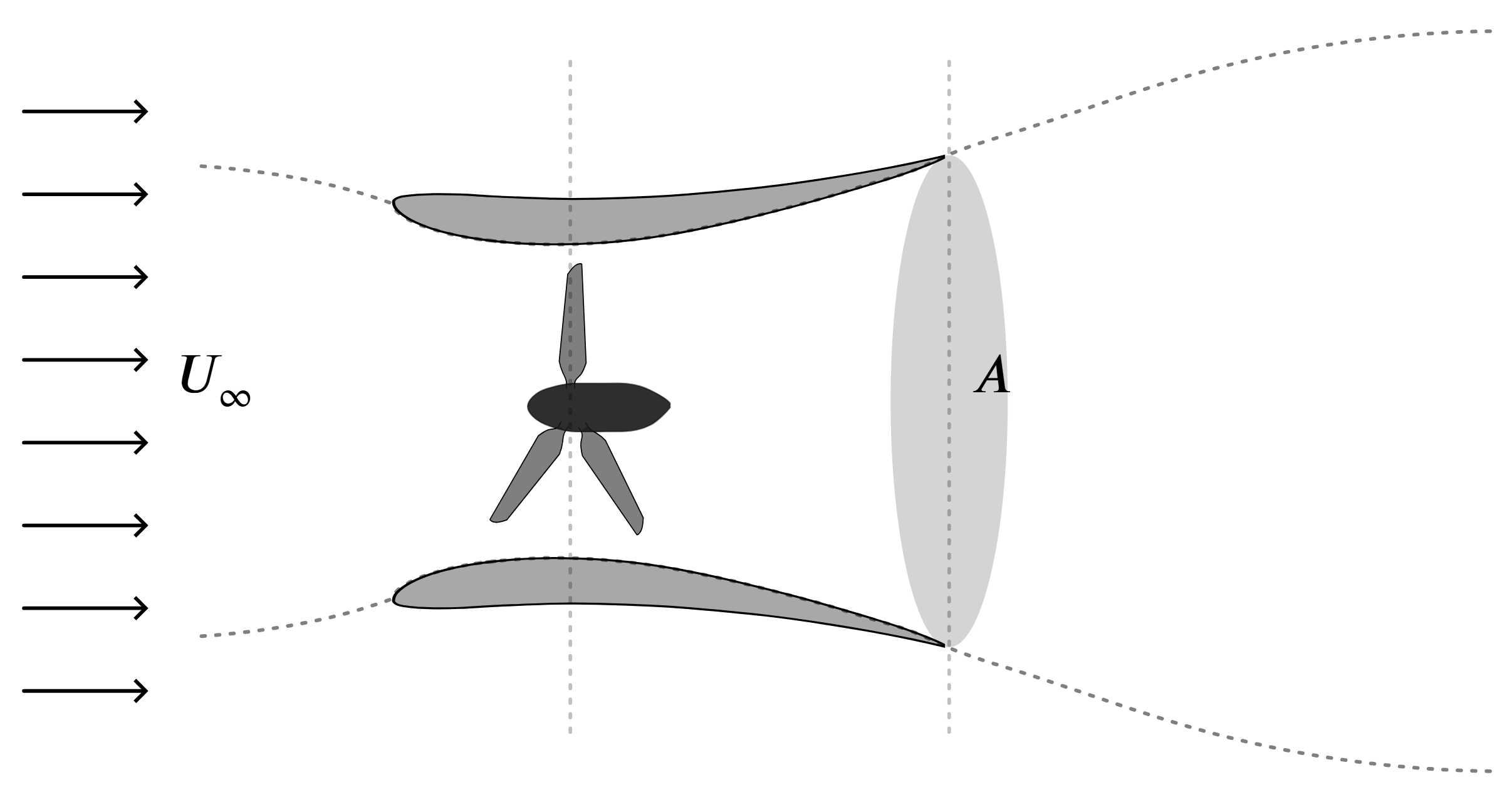}}
\caption{The ducted turbine has a projected area $A$ and is subject to inflow $U_{\infty}$.}
\label{fig:probstate}
\end{figure}

We consider a ducted hydrokinetic turbine operating in an unbounded fluid domain with uniform inflow $U_{\infty}$, as shown in Figure~\ref{fig:probstate}.
The available inflow power to the turbine is $P_{\text{avail}}=\rho A U^3_\infty/2$, where $\rho$ is the fluid density and $A$ is the area of the device perpendicular to the flow.
We consider $A$ as the maximum projection area of the duct, which in this case is at the duct exit.
The turbine converts the inflow power into rotational mechanical power with an efficiency characterized by the power coefficient
\begin{equation}
C_P=\frac{P}{P_{\text{avail}}},
\label{eq:CP}
\end{equation}
where $P=\tau\Omega$ is the mechanical power, with $\tau$ the turbine's torque and $\Omega$ its rotation speed.
The conversion of the turbine shaft mechanical power into electrical power through the generator system is subject to losses as follows:
\begin{equation}
P_e = P \cdot \eta, 
\label{eq:Pe}
\end{equation}
where $\eta$ is the electromechanical efficiency.
In this work, inflow velocity is set to $U_{\infty}=1.7~\text{m}/\text{s}$, based on data of the Mississippi River~\cite{USGS}, and $P_e=5~\text{kW}$.
The electrical components are sized based on this power and flow conditions~\cite{tariquzzaman2024multi}. 
This results in a generator with a diameter of $0.355~\text{m}$ and $\eta=0.8255$.
The device projection area $A$ is estimated by considering Eq.~\eqref{eq:CP} and \eqref{eq:Pe}, as well as $C_P=0.46$ according to the original Bahaj model~\cite{bahaj2007power}.
This yields $A=5.23~\text{m}^2$ associated with a diameter $D=2.58~\text{m}$.

Another two relevant non-dimensional parameters are the tip-speed ratio $\lambda$ and Reynolds number $Re$ (based on the diameter of the device), defined as
\begin{equation}
    \lambda = \frac{\Omega R}{U_{\infty}}, \ \ \ Re=\frac{U_{\infty}D}{\nu},
    \label{eq:TSR}
\end{equation}
where $\nu = 10^{-6}~\text{m}^2/\text{s}$ is the fluid kinematic viscosity. For the calculated $D$, we have $Re=4\times10^6$, and thus we consider the boundary layer to be fully turbulent.
The performance of the turbine will be evaluated as a function of $\lambda$ at this $Re$. 

The goal of the optimization is to simultaneously optimize the geometry of blades, hub, and duct to maximize $C_P$ at $U_\infty=1.7~\text{m}/\text{s}$ and fixed $\Omega=8.26~\text{rad}/\text{s}$, which corresponds to the optimal $\lambda$ of the baseline ducted turbine described later.
The designed turbine is expected to produce a power greater than $5~\text{kW}$ if the optimized $C_P$ is larger than 0.46 (the value used to calculate $A$).
If we need $P_e$ to be precisely $5~\text{kW}$, the size of the optimized device could be adjusted to achieve this value.
The complete optimization problem setup with 37 design variables is described in Table~\ref{table:optsetup}. 

\begin{table}[h!]
\centering
\begin{adjustbox}{width=0.8\columnwidth,center}
\begin{tabular}{llll}
\hline
    & \textbf{Variables/Function} & \textbf{Description} & \textbf{Quantity} \\
    \hline
    & & & \\
    \textbf{Maximize}        & $C_P$ & Power coefficient & 1 \\   
    & & & \\
    \textbf{with respect to} & $\alpha$ & Duct cone angle & 1 \\
                             & $A_{\text{upper}}$ & Duct upper polynomial coefficients & 6 \\
                             & $A_{\text{lower}}$ & Duct lower polynomial coefficients & 6 \\
                             & $\beta$ & Duct length & 1 \\ 
                             & $R_b$ & Blade radius & 1 \\
                             & $\theta$ & Blade twists & 9 \\
                             & $c$ & Blade chords & 9 \\ 
                             & $x_{\text{hub}}$ & Hub control point translations & 4 \\
    \cline{3-4}
    & & Total & 37 \\
    & & & \\
    \textbf{subject to}      & $t_{\text{duct}}\geq 0.014~\text{m}$ & Duct thickness & 50 \\
                             & $h=0.05~\text{m}$ & Tip gap & 1 \\
                             & $\rho_{\kappa} \geq 0.0015~\text{m}$ & Duct leading edge curvature radii & 1 \\
                             & $A=5.23~\text{m}^2$ & Maximum area & 1 \\
                             & $D_{\text{hub}}=0.4~\text{m}$ & Hub cylindrical section diameter & 1 \\
                             & $L_{\text{hub}}=0.78~\text{m}$ & Hub cylindrical section length & 1 \\
    \cline{3-4}
    & & Total & 55 \\
\hline\hline
\end{tabular}
\end{adjustbox}
\caption{Optimization problem statement}
\label{table:optsetup}
\end{table}

\begin{figure}[h!]
\centering
\begin{subfigure}[t]{0.48\textwidth}
    \centering
    \includegraphics[height=5cm]{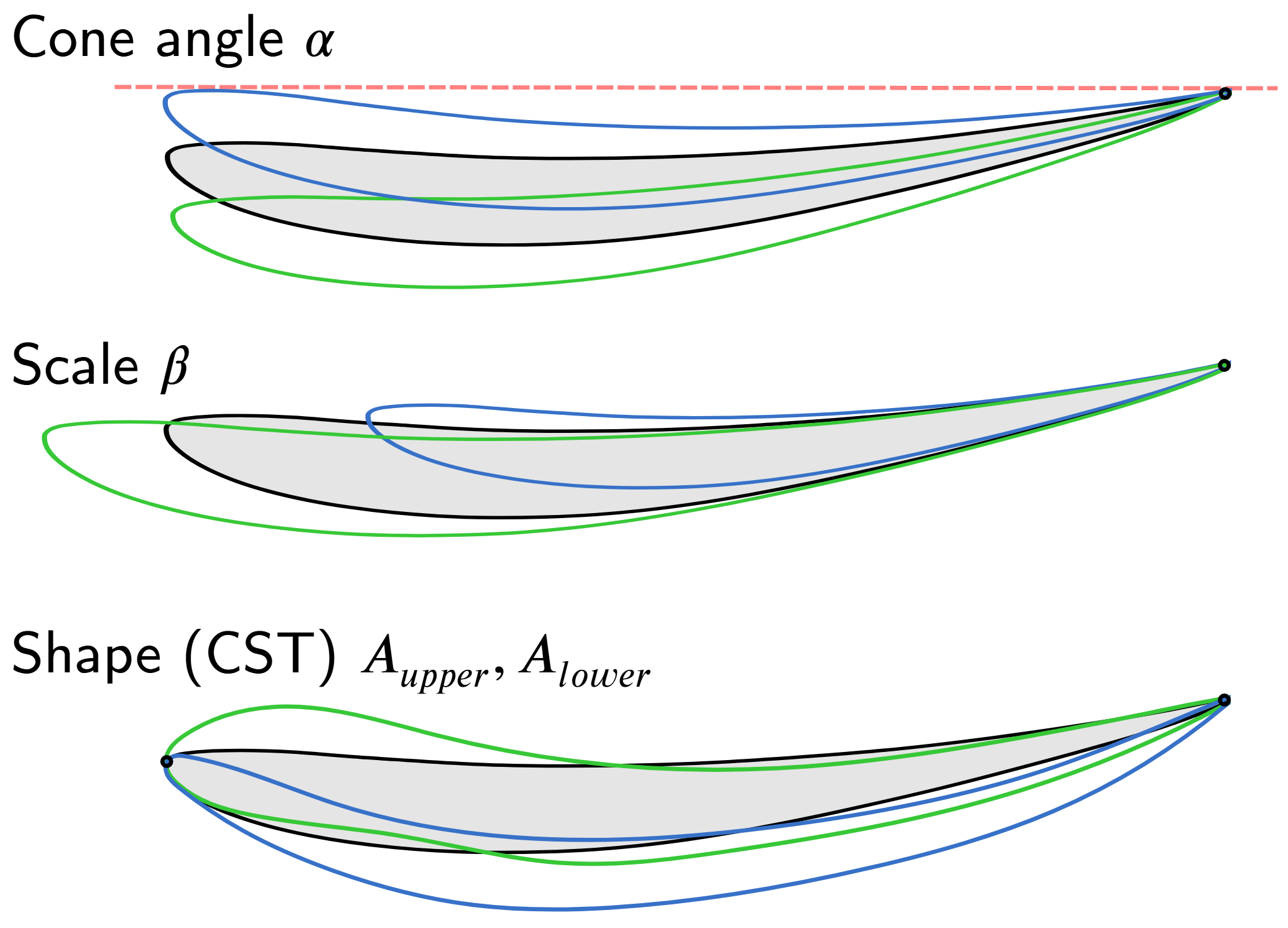}
    \caption{Duct design variables}
    \label{fig:optprob:ductDV}
\end{subfigure}%
\hfill
\begin{subfigure}[t]{0.48\textwidth}
    \centering
    \includegraphics[height=5cm]{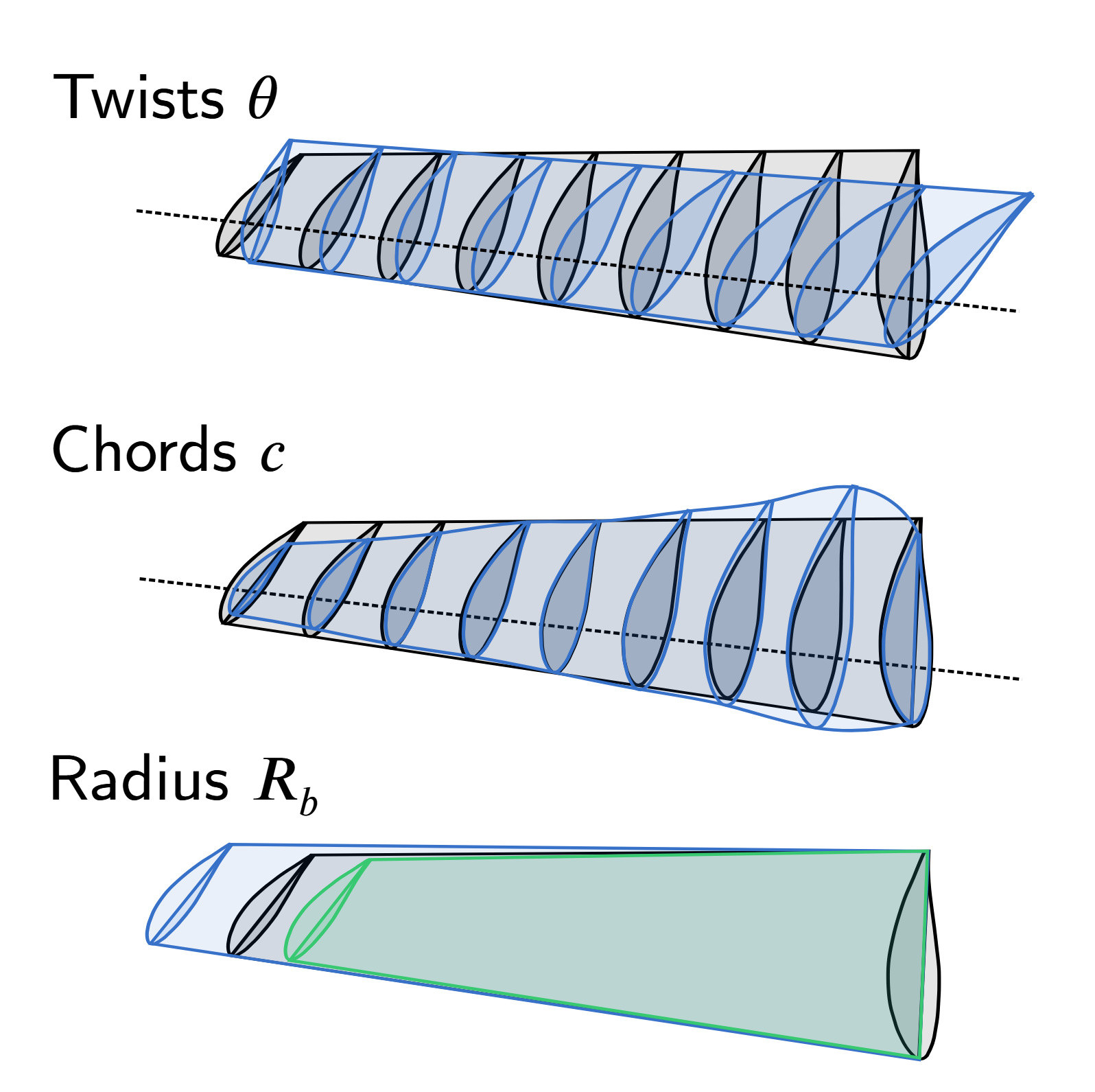}
    \caption{Blade design variables}
    \label{fig:optprob:bladeDV}
\end{subfigure}
\vspace{0.5cm} 
\begin{subfigure}[t]{0.48\textwidth}
    \centering
    \includegraphics[width=\linewidth]{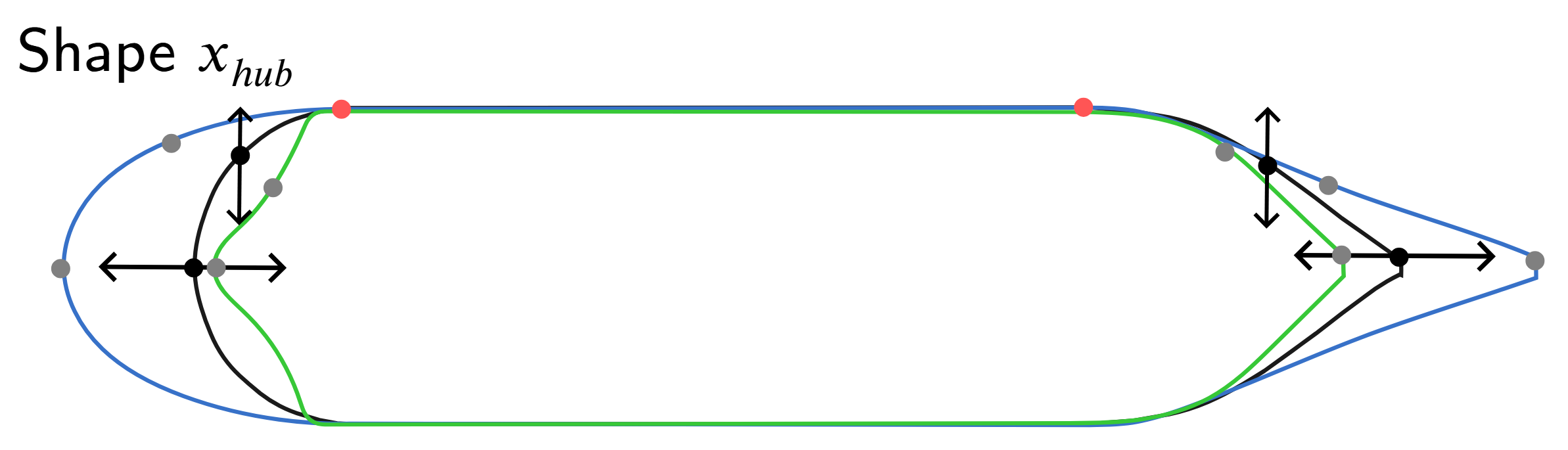}
    \caption{Hub design variables}
    \label{fig:optprob:hubDV}
\end{subfigure}
\caption{Design variables for (a) duct, (b) blade, and (c) hub. The baseline design is shown in black, with blue and green illustrating two examples of deformed configurations.}
\label{fig:optprob:DVs}
\end{figure}

The duct design is parameterized by 14 variables that control the duct cone angle, length, and upper and lower surface shapes, as illustrated in Figure~\ref{fig:optprob:ductDV}. 
The duct cone angle is adjusted by pivoting around the trailing edge, with a bound on maximum rotation to ensure that no sectional area exceeds the exit area, $A=5.23~\text{m}^2$.
The duct scaling is also anchored to the trailing edge, with a scaling factor applied to all points on the duct.
Selecting the trailing edge as the reference point makes it easier to keep the exit diameter as the maximum.
The upper and lower surfaces of the duct are parameterized using class-shape transformation (CST) variables~\cite{kulfan2006fundamental}, which are coefficients of the Bernstein basis polynomials as detailed in Appendix~\ref{app:cst}.
The CST parametrization enables broad design exploration while maintaining a smooth foil-shaped cross-section.

The turbine blades are parameterized by 19 variables, including the chords and twists at 9 spanwise locations and the blade radius, as illustrated in Figure~\ref{fig:optprob:bladeDV}.
The chords are determined by scaling factors that change the size of the section while maintaining the sectional foil shape, e.g., the thickness-to-chord ratio. 
Any changes to these 19 variables are equally applied to all three blades.
The turbine hub is parameterized by translating 4 control points.
Two points at the front and rear hub ends move axially, primarily driving the hub length.
The other two points control the hub front and rear curvature by moving in the radial direction, as shown in figure~\ref{fig:optprob:hubDV}.
Once the locations of the 4 points are given, two cubic splines are used to fit the front and rear parts of the hub.
We ensure slope continuity at the front end and the connection to the main hub cylindrical surface.
This cylindrical section is fixed at a diameter $D_{\text{hub}}=0.4~\text{m}$ and length $L_{\text{hub}}=0.78~\text{m}$ to accommodate the selected generator inside.

In addition to the constraints mentioned above on $A$, $D_{\text{hub}}$, and $L_{\text{hub}}$, we also impose a constraint on minimum duct thickness $t_{\text{duct}}=0.014~\text{m}$ to meet the yield strength requirements (based on a calculation using 6061 aluminum alloy).
This constraint is enforced at 50 chordwise duct sections for global satisfaction.
The duct leading-edge radius of curvature is constrained to a minimum of $\rho_{\kappa}=0.0015~\text{m}$. 
This minimum leading-edge radius improves off-design performance at oblique flow conditions~\cite{Madsen2019a,Mangano2021a}.
Finally, a constraint on a constant tip gap of $h=0.05~\text{m}$ is set between the duct and blade surfaces.
In the optimization, the constraint values $A$, $t_{\text{duct}}$, $\rho_{\kappa}$, and $h$ can all be computed as direct or indirect functions of the design variables (see Appendix~\ref{app:cst}, particularly for $\rho_{\kappa}$) and can therefore be treated as standard equality or inequality constraints.

In this work, the baseline blade geometry is taken from the optimized design A in our previous work~\cite{park2023cfd}, with NACA~63-8xx sectional shapes and a blade radius of $1~\text{m}$.
The baseline hub geometry consists of a hemisphere for the front part and a half ellipsoid of length $0.527~\text{m}$ for the rear part.
The baseline duct uses an Eppler~423 (E423) foil shape, which is known for its high lift coefficient~\cite{venters2018ducted}, scaled for a chord length of $3.013~\text{m}$ and maximum thickness of $0.129~\text{m}$ and then rotated to maintain the tip gap of $h=0.05~\text{m}$ --- see Figure~\ref{fig:results:optDuctHub} and \ref{fig:results:optBlade} in Section~\ref{subsec:results:opt}.

\section{Method}
\label{sec:Method}

\begin{figure}[h!]
\centerline{\includegraphics[width=\columnwidth]{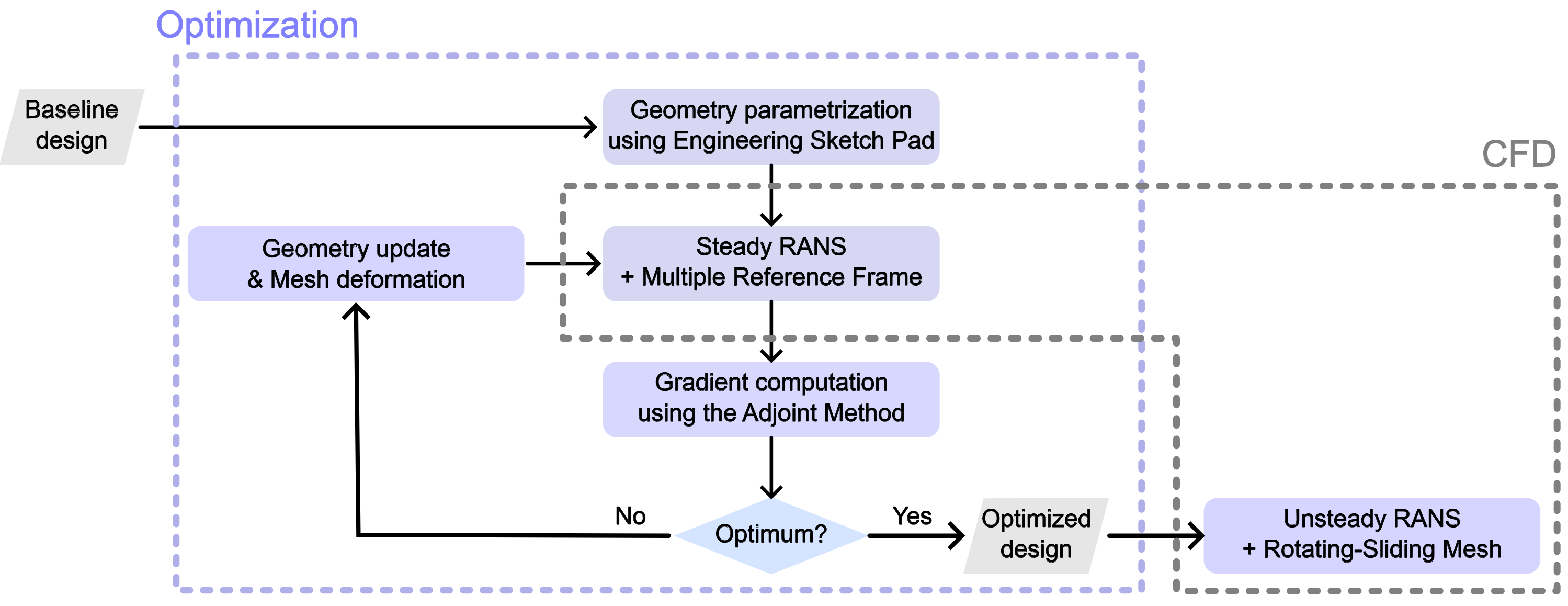}}
\caption{Flowchart of the optimization and CFD simulation processes.}
\label{fig:flowchart}
\end{figure}

Our overall methodology is summarized in Figure~\ref{fig:flowchart}, which shows two major components intertwined with each other: gradient-based optimization and CFD simulations.
The key procedure is described as follows.
\begin{enumerate}
    \item The baseline geometry is parameterized with Engineering Sketch Pad (ESP) method implemented in the pyGeo package~\cite{Kenway2010b,Hajdik2023c}. A  computational mesh is generated using Pointwise~\footnote{Cadence Fidelity Pointwise for CFD Meshing \url{https://www.cadence.com/en_US/home/tools/system-analysis/computational-fluid-dynamics/fidelity.html\#fidelity-pointwise}}.
    \item DAFoam's RANS solver with multiple reference frames (MRF) and Spalart--Allmaras (SA) turbulence model~\cite{Spalart1992a} evaluates the performance of the ducted turbine and predicts a $C_P$.
    \item The gradient $dC_P/d\bm{x}$ is computed using the discrete adjoint method, where $\bm{x}\in \mathbb{R}^{37}$ is the vector of design variables.  The discrete adjoint method is an approach for efficient derivative computation, with the computational cost independent of the dimension of $\bm{x}$. This solver is also implemented in DAFoam.
    \item The next design point is identified through a sequential quadratic programming algorithm implemented in SNOPT~\cite{Gill2005a}, using information from Steps~2 and 3, as well as the specified constraints.
    \item The computational mesh is deformed according to the new set of design variables $\bm{x}$ in two steps. First, the surface mesh is updated based on the updated ESP geometry. Then, the surface deformation is propagated to the volume mesh through the analytic inverse-distance weighting method~\cite{Luke2012a} implemented in the IDWarp package~\cite{Secco2021a}.
    \item Steps 2 to 5 are repeated until the optimality convergence criteria are met, after which the optimized design is obtained.
    \item URANS solver with a rotating-sliding mesh (RS) in OpenFOAM~\cite{jasak2007openfoam} and $k-\omega$ shear stress transport (SST) turbulence model~\cite{menter2003ten} re-evaluates the performance of the optimized design on a mesh generated using \emph{snappyHexMesh}.
\end{enumerate}

Most components involved in the above procedure are detailed in our previous work~\cite{park2023cfd}, except for the ESP parametrization and its integration into the mesh generation and deformation within the optimization process.
In the following sections, we provide detailed descriptions of the ESP parametrization and the associated mesh generation using Pointwise, followed by the mesh generated with \emph{snappyHexMesh} for URANS re-evaluation in Step~7.
The key files for this work are publicly available~\footnote{\url{https://github.com/jbpark94/Ducted-Hydrokinetic-Turbine-Key-Files.git}}.

\subsection{ESP and Pointwise mesh}
\label{subsec:Method:esp}

Our previous optimization work~\cite{park2023cfd} had two major limitations: the exclusion of a hub and the restriction of a duct to a thin-walled cross-section, as discussed in Section~\ref{sec:Intro}. 
These limitations were related to the FFD approach~\cite{Sederberg1986,Kenway2010b} that we used for the geometry parametrization.
In particular, it is challenging to separately control the deformation of the hub and blades through FFD, as they are attached and have to be controlled by the same FFD box enclosing the geometry. Moreover, when FFD manipulation is applied to a foil-shaped duct, the deformation inevitably breaks the circular and axisymmetric shape.
These limitations become particularly problematic as multiple tightly spaced components require independent and precise control. To overcome these limitations, we use ESP~\cite{Haimes2013a} in this work.

\begin{figure}[h!]
\centering
\begin{subfigure}[t]{0.44\textwidth}
    \centering
    \includegraphics[width=\textwidth]{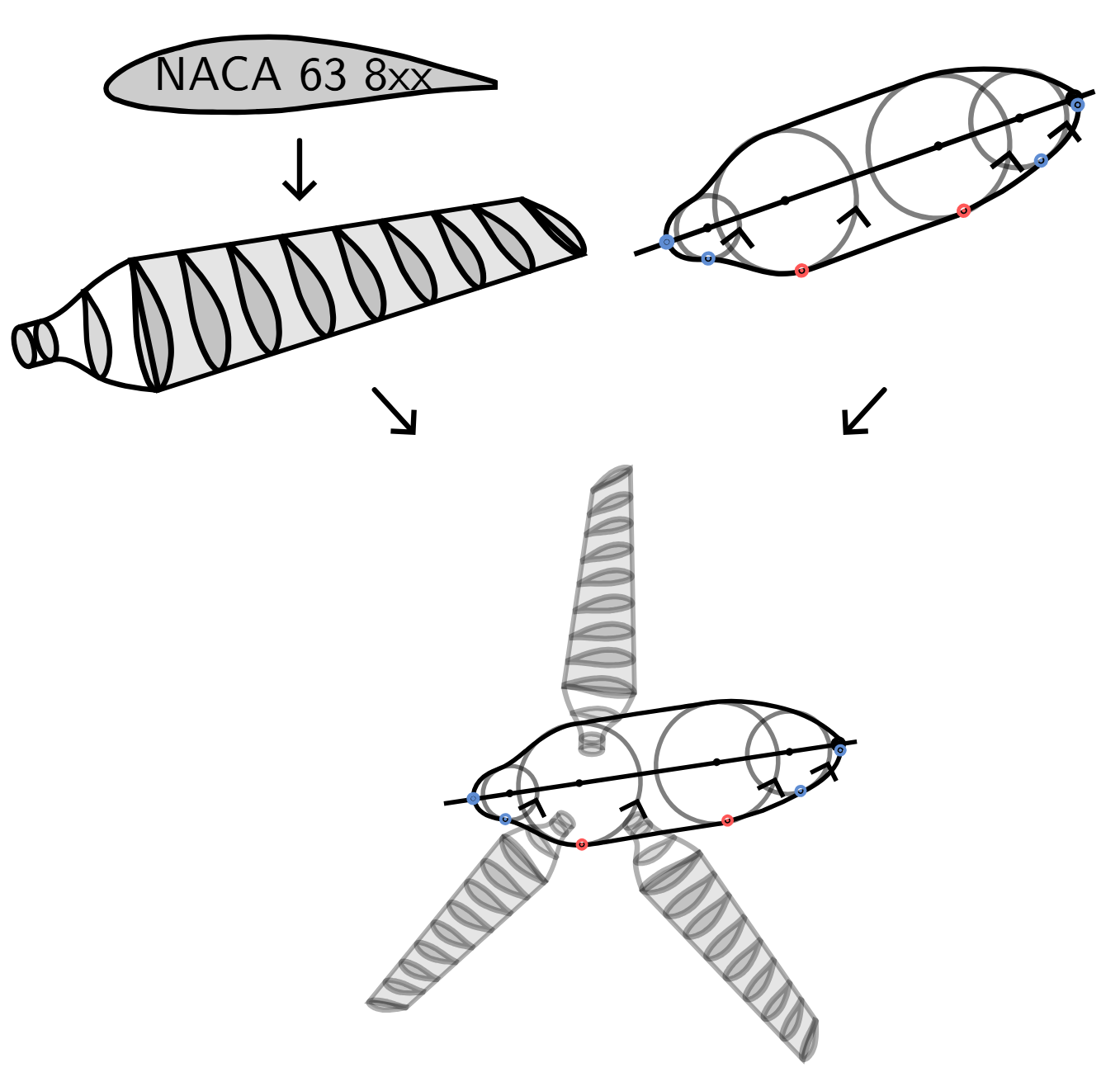}
    \caption{ESP model of the turbine.}
    \label{fig:optprob:turbine}
\end{subfigure}%
\hfill
\begin{subfigure}[t]{0.44\textwidth}
    \centering
    \includegraphics[width=\textwidth]{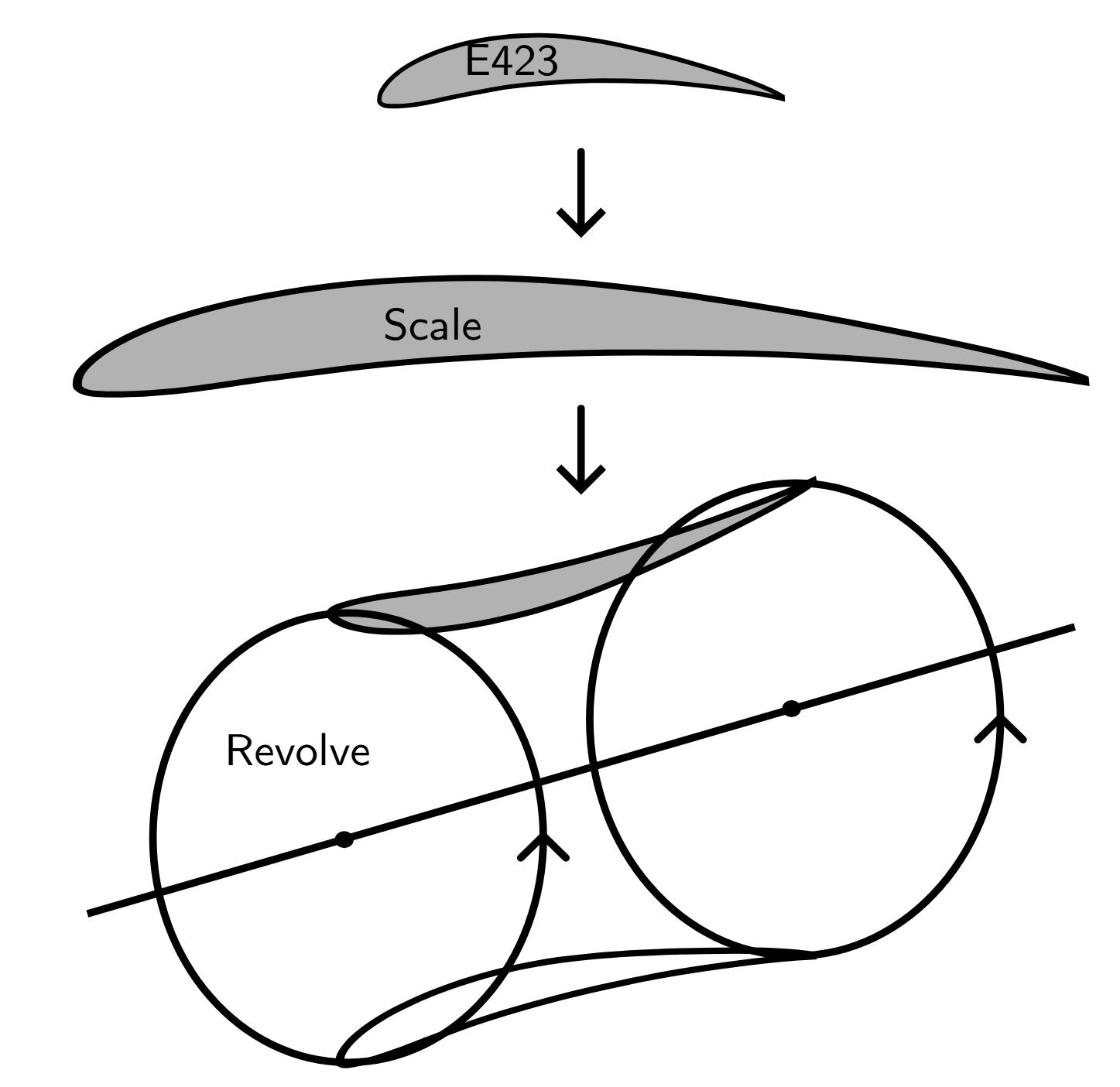}
    \caption{ESP model of the duct.}
    \label{fig:optprob:duct}
\end{subfigure}
\caption{ESP model of the ducted turbine. (a) The turbine geometry is created by combining three blades and a hub. The blades are generated by blending NACA~63-8xx hydrofoils along the span, with rotation and scaling applied according to the desired twist and chord distributions.
The hub shape is generated by spline curves based on four control points, ensuring slope continuity at the hub's front end and the junctions with the central cylindrical section and revolving the curves around the hub's axial axis. (b) The duct cross-section is defined using 12 CST variables, a conning angle, and a scale factor. The cross-section revolves around the axis to create the 3D duct geometry.}
\label{fig:ESP:geo}
\end{figure}

The entire ducted turbine geometry is generated and parameterized using ESP, with the design variables listed in Table~\ref{table:optsetup} and illustrated in Figure~\ref{fig:ESP:geo}. 
ESP generates the blade surface by blending sectional NACA~63-8xx shapes at nine spanwise locations with desired twists $\theta$ and sectional chords $c$ using cubic spline surfaces.
For the hub, cubic splines are generated using four control points to shape the front and rear sections, ensuring slope continuity at the hub's front end and junctions with the central cylindrical section of the hub.
The curves are then revolved around the hub's longitudinal axis to form the full axisymmetric hub surface. 
The duct foil shape is defined by the CST variables $A_{upper}$ and $A_{lower}$ scaled by a factor $\beta$ and rotated by an angle $\alpha$ with respect to the trailing edge, which is then revolved around the axial axis to create the complete duct geometry. 

\begin{figure}[h!]
\centering
\begin{subfigure}[t]{0.48\textwidth}
    \centering
    \includegraphics[width=\textwidth]{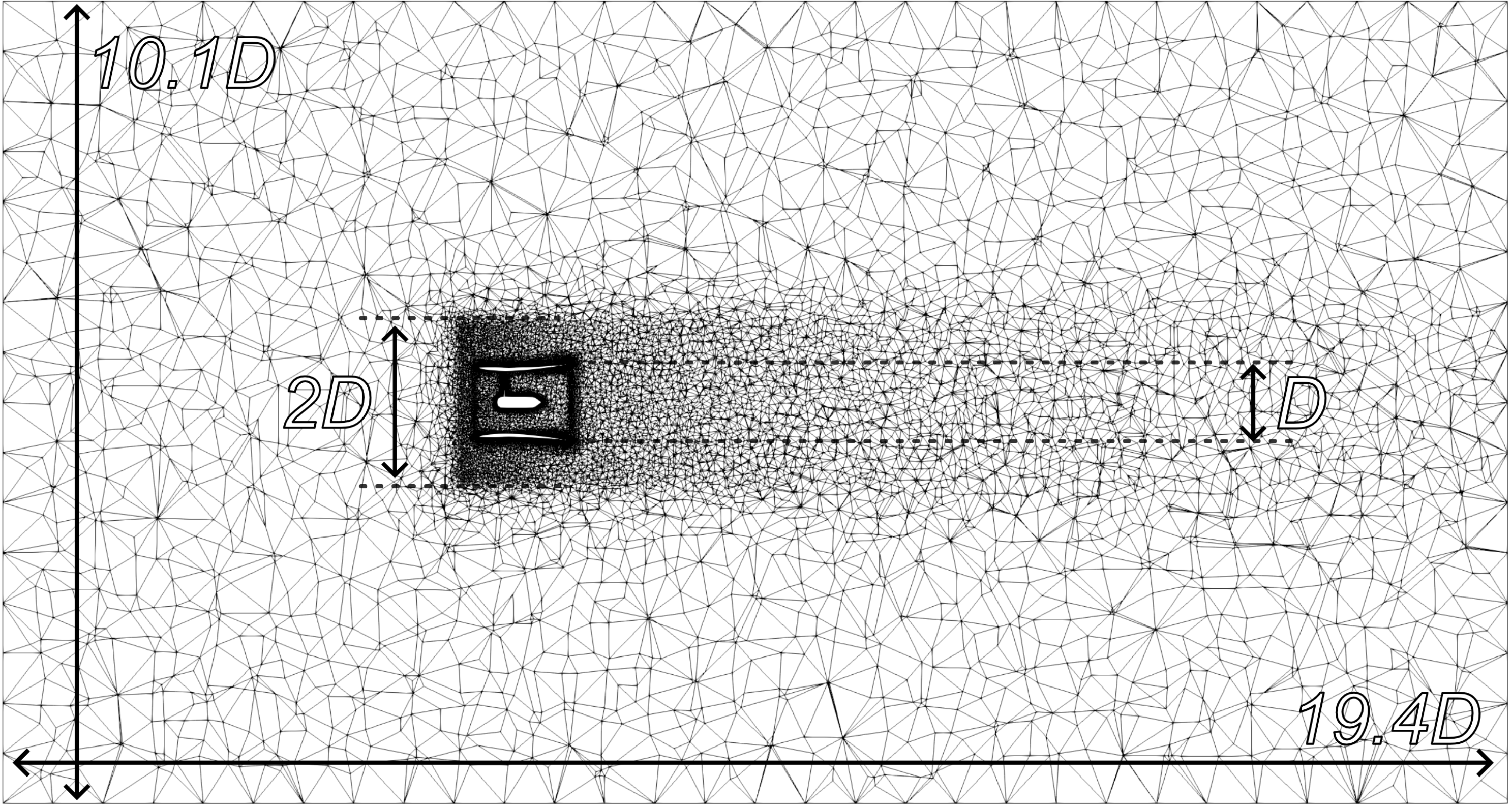}
    \caption{Side view}
    \label{fig:pw1}
\end{subfigure}%
\hfill
\begin{subfigure}[t]{0.48\textwidth}
    \centering
    \includegraphics[width=\textwidth]{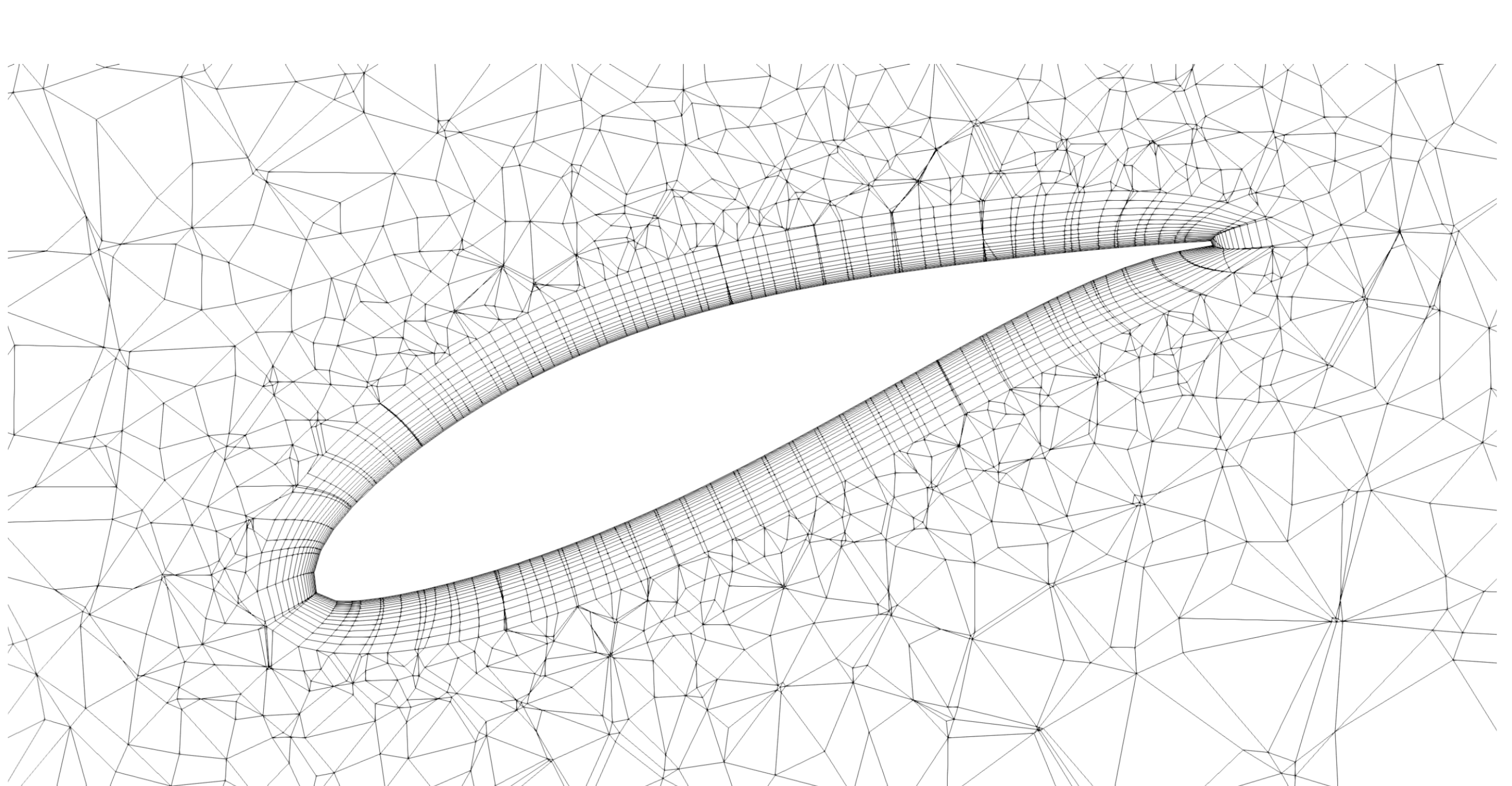}
    \caption{Close-up of the $0.5R$ blade section}
    \label{fig:pw2}
\end{subfigure}
\begin{subfigure}[t]{0.48\textwidth}
    \centering
    \includegraphics[width=\textwidth]{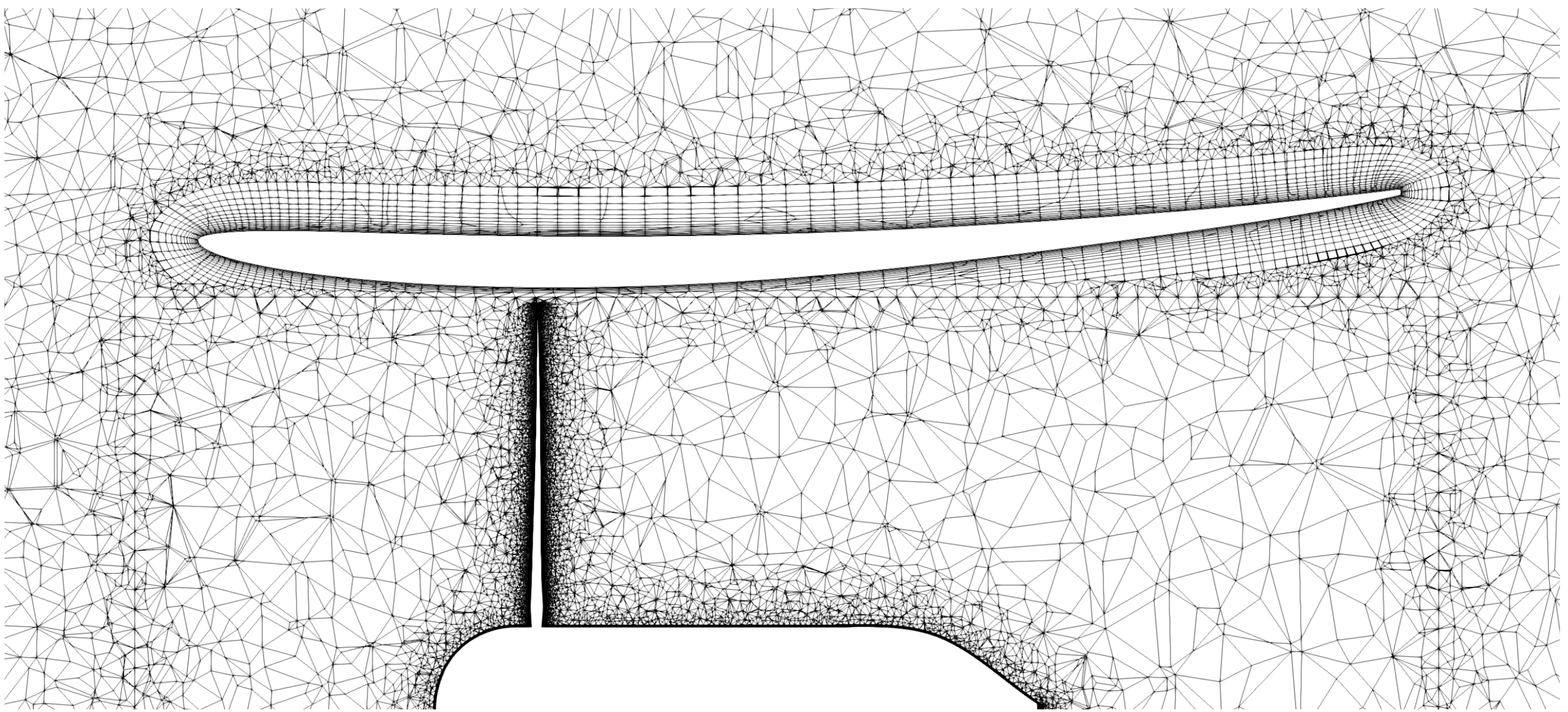}
    \caption{Close-up near the duct and hub}
    \label{fig:pw3}
\end{subfigure}
\caption{The mesh generated by using Pointwise predominantly composed of tetrahedral cells. The domain of the mesh is $10.1D\times10.1D\times19.4D$. 
The mesh is refined around the ducted turbine geometry, and the near-surface regions are further refined using 15 prism layers with an expansion ratio $1.2$ and an initial cell spacing of $0.001$~m for the duct and $5\times 10^{-5}$~m for the blade and hub, resulting in a total cell count of 6.3 million.}
\label{fig:pw}
\end{figure}

We use Pointwise to generate high-quality surface and volume meshes.
Based on the geometry information provided by ESP, we discretize the surfaces using triangular elements for the blade and hub surfaces and quadrilateral elements for the duct surface. 
The volume mesh, mainly composed of tetrahedral elements, is then generated to fill the space between the body surface and the domain boundary.
Our computational mesh has a total of 6.3 million cells and is shown in Figure~\ref{fig:pw}.
Meshing with Pointwise allows a precise representation of the body surfaces, which is difficult to achieve with other automated methods such as \emph{snappyHexMesh}, used in our previous work~\cite{park2023cfd}.
Inaccurate surface representations are usually more problematic at corners such as the trailing edges of the blades, which can lead to poor optimization convergence, among other issues. 

The mesh deformation in Step~5 also requires ESP integration.
To understand this, let us first express the mapping from the ESP surface to the mesh nodes as a function $f_S$.
In particular, for an ESP-generated cubic spline surface $S$, the location of any mesh node on it is written as
\begin{equation}
    \bm{X}_i = f_S(\bm{x},u_i,v_i),
\label{eq:param}
\end{equation}
where $u_i\in[0,1]$ and $v_i\in[0,1]$ are normalized local coordinates of the surface $S$, $\bm{X}_i$ is the nodal location in global coordinate with $i$ looping over all nodes on surface $S$.
In a surface mesh, the locations of $u_i$ and $v_i$ are stored for all nodal locations.
When the design variable $\bm{x}$ is updated, the new location of $\bm{X}_i$ is computed with $u_i$, $v_i$, and the function $f_S$ remaining the same~\cite{Brelje2021}.
The volume mesh is then updated using the analytic inverse-distance weighting method~\cite{Luke2012a,Secco2021a}.

\subsection{\emph{snappyHexMesh} for URANS simulations}
\label{subsec:Method:snappy}

\begin{figure}[h!]
\centering
\begin{subfigure}[t]{0.49\textwidth}
    \centering
    \includegraphics[width=\textwidth]{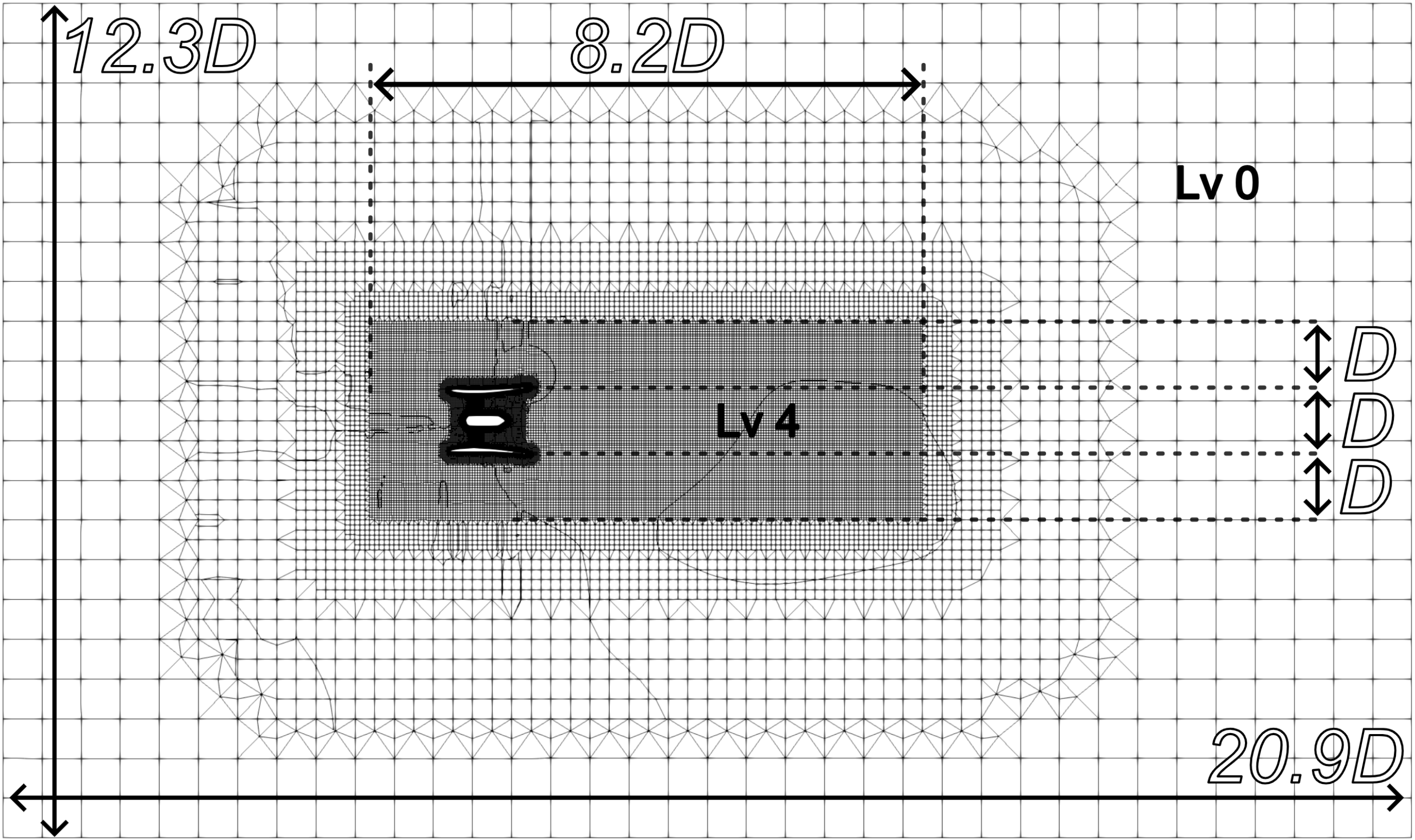}
    \caption{Side view}
    \label{fig:snappy1}
\end{subfigure}%
\hfill
\begin{subfigure}[t]{0.49\textwidth}
    \centering
    \includegraphics[width=\textwidth]{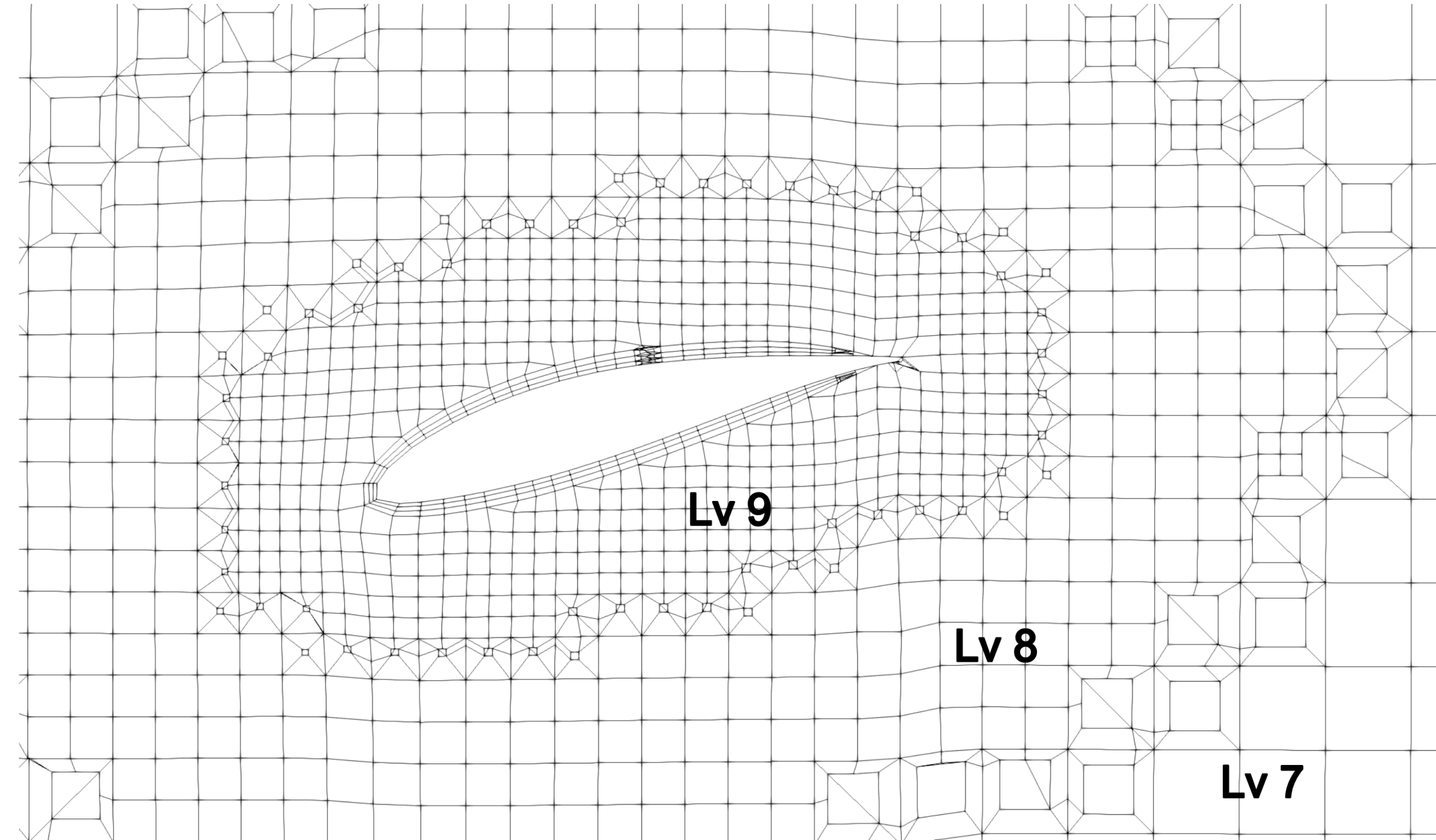}
    \caption{Close-up of the blade section at $0.5R$}
    \label{fig:snappy2}
\end{subfigure}
\begin{subfigure}[t]{0.48\textwidth}
    \centering
    \includegraphics[width=\linewidth]{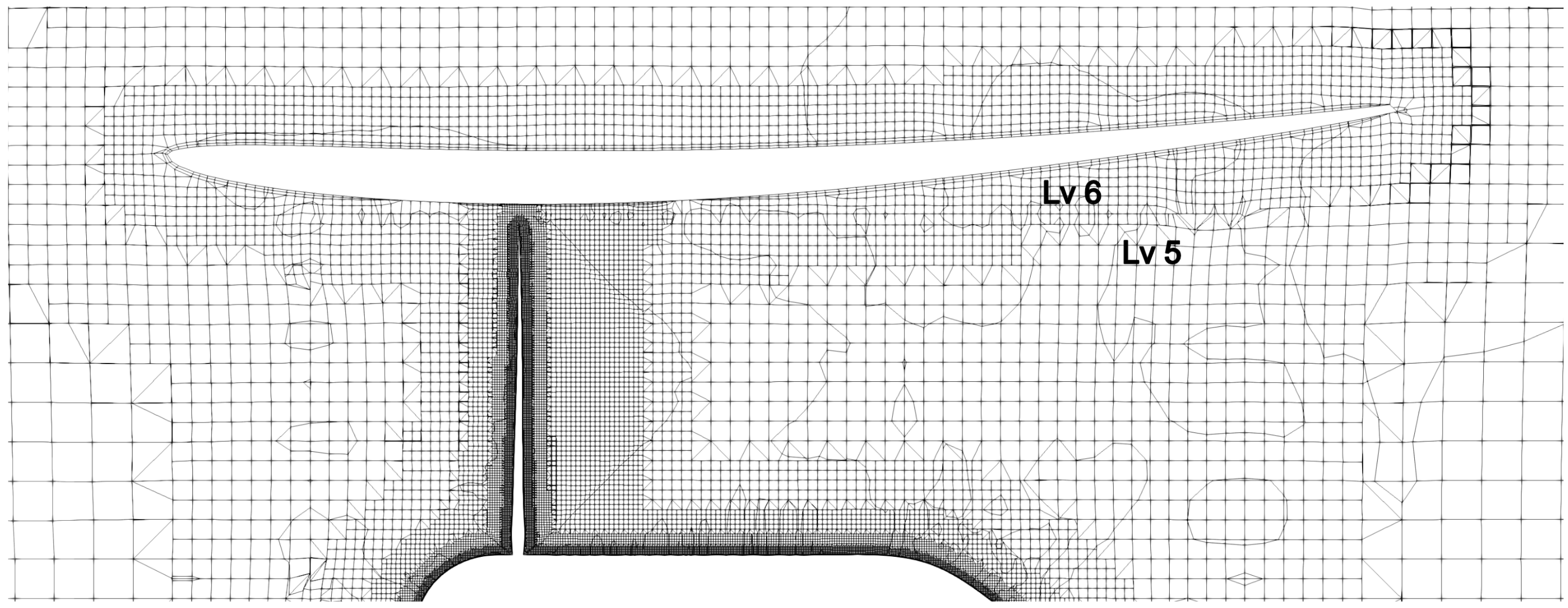}
    \caption{Close-up near the duct and hub}
    \label{fig:snappy3}
\end{subfigure}
\caption{The mesh generated using \emph{snappyHexMesh} predominantly contains hexahedral cells. The domain of the mesh is $12.3D\times12.3D\times20.9D$. We add a refinement region around the ducted turbine with a size of $3D\times3D\times8.2D$, with prism layers that further resolve the near-wall region.} 
\label{fig:snappy}
\end{figure}

We switch to \emph{snappyHexMesh} for the higher-fidelity URANS-RS re-evaluations in Step~7.
While the Pointwise mesh provides the exact representation of the body surfaces, it usually includes highly non-orthogonal cells.
These cells prohibit efficient URANS simulations because additional correction steps are needed at each iteration, and the whole process is costly in our current setting.
The mesh generated using \emph{snappyHexMesh} is predominantly composed of hexahedral cells with a domain of size $12.3D\times12.3D\times20.9D$, as shown in Figure~\ref{fig:snappy}.
In a region of $3D\times3D\times8.2D$ surrounding the turbine, each cell undergoes a level-4 refinement, subdividing each initial cell into $2^4$ cells per direction.
Further refinements of levels 6 and 9 are applied toward the duct, blade, and hub surfaces, respectively.
Three prism layers with an expansion ratio of 1.2 are added near the turbine and duct surfaces to improve the near-wall resolution.
We use these strategies to generate meshes with different cell counts by varying the resolution of the base mesh.
Ducted turbine performance will be evaluated on these different meshes.

\section{Results and Discussion}
\label{sec:Results}

\begin{figure}[h!]
\centerline{\includegraphics[width=\textwidth]{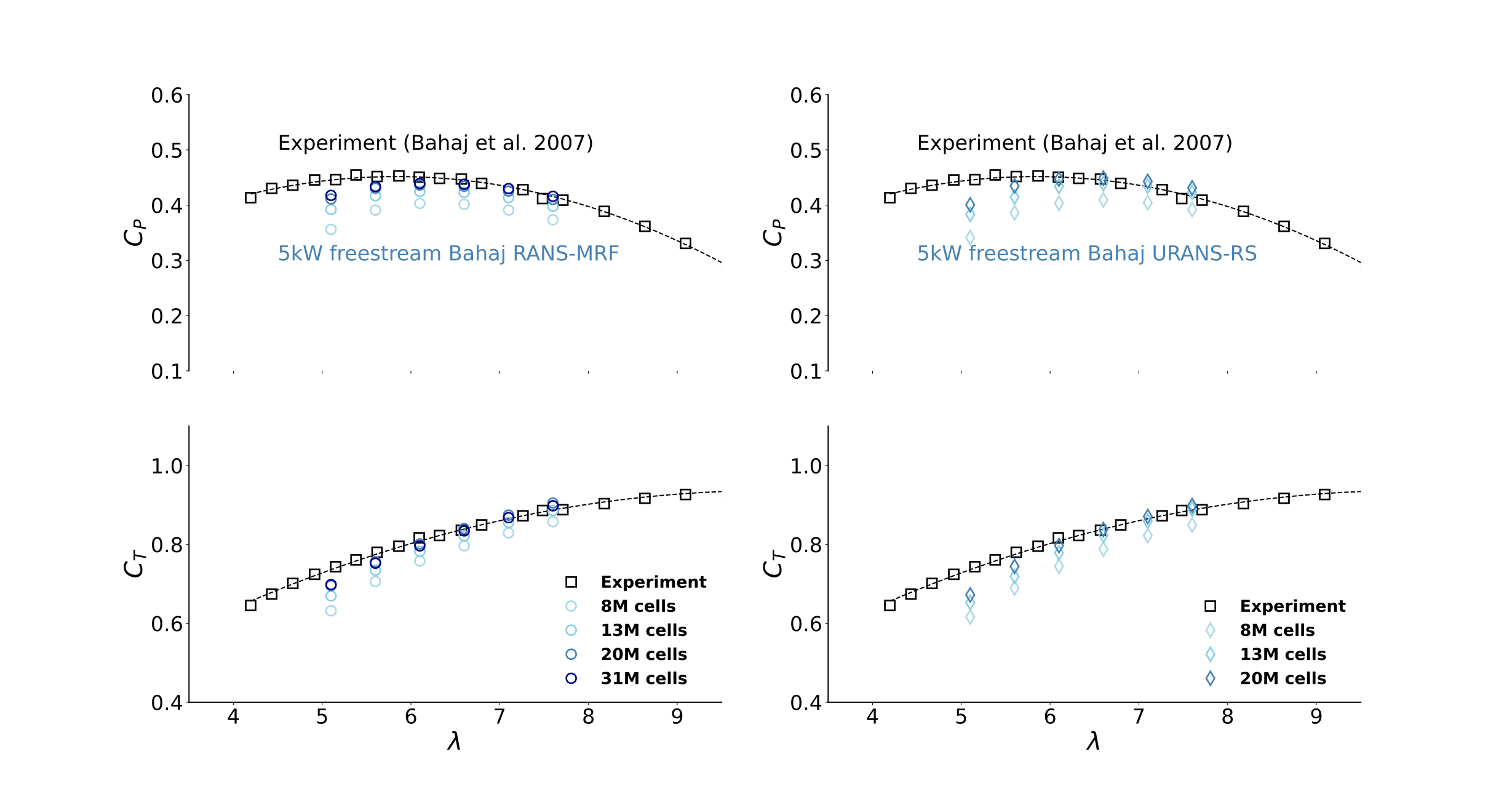}}
\caption{Comparison between RANS (left column) and URANS solvers (right column) with varying mesh resolutions against experimental results for the Bahaj turbine~\cite{bahaj2007power}.}
\label{fig:results:bahaj}
\end{figure}

Before presenting the optimization results, we first present the RANS and URANS analysis for a $5~\text{kW}$ freestream Bahaj turbine featuring the baseline hub, compared to the experimental data at a smaller model scale~\cite{bahaj2007power}.
Figure~\ref{fig:results:bahaj} plots the power coefficient $C_P$ and thrust coefficient $C_T=T/(0.5\rho U_{\infty}^2 A)$ for various mesh resolutions. 
For the RANS simulations, we consider the results converged when the residual of the momentum equation decreases below $\mathcal{O}(10^{-6})$.
For the URANS simulations, we simulate approximately 13 rotations of the turbine ($10~\text{s}$ in simulated physical time) and calculate the turbine performance in the stationary state (which is reached after about $5~\text{s}$).
Both the RANS-MRF and URANS-RS results converge well with mesh refinement and compare reasonably with the experimental data.
When interpreting the difference between simulation and experimental results, we need to keep in mind that the simulations are performed for a much larger turbine (of blade radius $1.29~\text{m}$, approximately three times larger than the experiments) and a hub with a much larger relative size (the exact size not specified for experiments in the paper).

\begin{figure}[h!]
\centerline{\includegraphics[width=\textwidth]{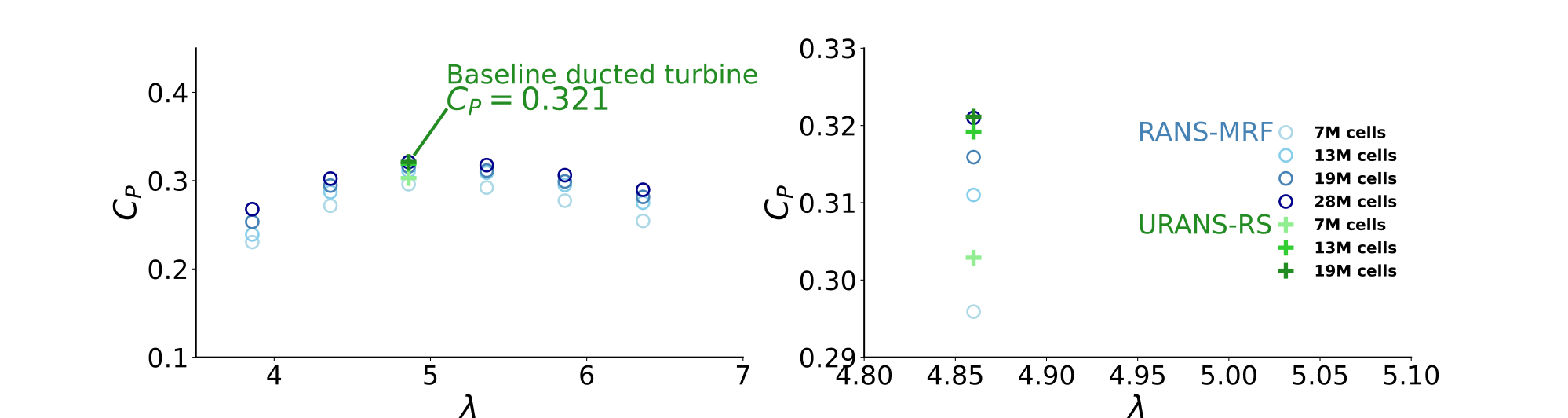}}
\caption{Power coefficient $C_P$ of the baseline ducted turbine as a function of $\lambda$ evaluated using RANS-MRF  (\textcolor[HTML]{4682b4}{\boldmath$\mathbf{\circ}$}) and URANS-RS (\textcolor[HTML]{228b22}{\boldmath$\mathbf{+}$}) with varying mesh resolutions. The right panel provides a close-up view of the results at $\lambda=4.86$.}
\label{fig:base}
\end{figure}

Next, we evaluate the power coefficient $C_P$ of the baseline ducted turbine design as a function of tip speed ratio $\lambda$ using the RANS-MRF approach.
As shown in Figure~\ref{fig:base}, the peak $C_P=0.321$ is obtained at $\lambda=4.86$ (corresponding to a turbine rotation rate of $\Omega=8.26~\text{rad}/\text{s}$) among 6 values of $\lambda$ ranging from $3.86$ to $6.36$.
At $\lambda=4.86$, we re-evaluate the performance using the URANS-RS method, which provides a similar value of $C_P$.
The optimization problem described in Table~\ref{table:optsetup} is then performed at $\lambda=4.86$.

\subsection{Optimization}
\label{subsec:results:opt}

\begin{figure}[h!]
\centerline{\includegraphics[width=0.75\textwidth]{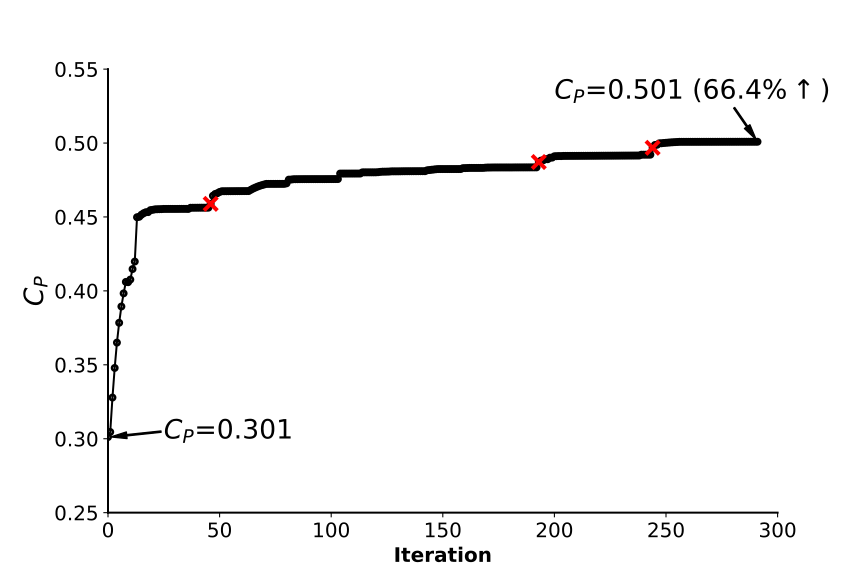}}
\caption{Power coefficient $C_P$ versus iteration during the optimization process. Red crosses indicate the points where re-meshing becomes necessary due to significant mesh distortion.}
\label{fig:results:iter}
\end{figure}

We solve the optimization problem described in Section~\ref{sec:Probstate} following the procedure outlined in Section~\ref{sec:Method}.
Figure~\ref{fig:results:iter} shows the evolution of the power coefficient $C_P$ throughout the optimization process.
Starting from the baseline design with $C_P=0.301$, the optimization achieves a final design with $C_P=0.501$ after 291 iterations.
This took 17 days and 2 hours of wall time using 512 cores in Intel Xeon 8352Y processors on the TAMU FASTER cluster.
The initial $C_P$ differs from the value $0.321$ shown in Figure~\ref{fig:base} because of the coarser mesh, the SA turbulence model, and a lower-order numerical scheme in optimization for stable adjoint solver convergence.
This is similar to what we found in our previous work~\cite{park2023cfd}.
The red crosses on the graph indicate points where re-meshing is necessary due to excessive mesh distortion that triggers IDWarp failures.

The optimization is considered complete when $C_P$ plateaus, despite further mesh adjustments.
While this approach is not numerically rigorous, it is appropriate for practical engineering design considerations.
The SNOPT optimality metric~\cite{Gill2005a} decreases from $10^{-1}$ to $10^{-2.28}$, which is a reduction of more than one order of magnitude.
Numerical noise on the function derivatives and the re-meshing process prevents a lower optimality value.
When $C_P$ remains unchanged with additional re-meshing, we treat this as the optimal point where further $C_P$ gains would likely be overshadowed by numerical noise.

\begin{figure}[h!]
\centerline{\includegraphics[width=\textwidth]{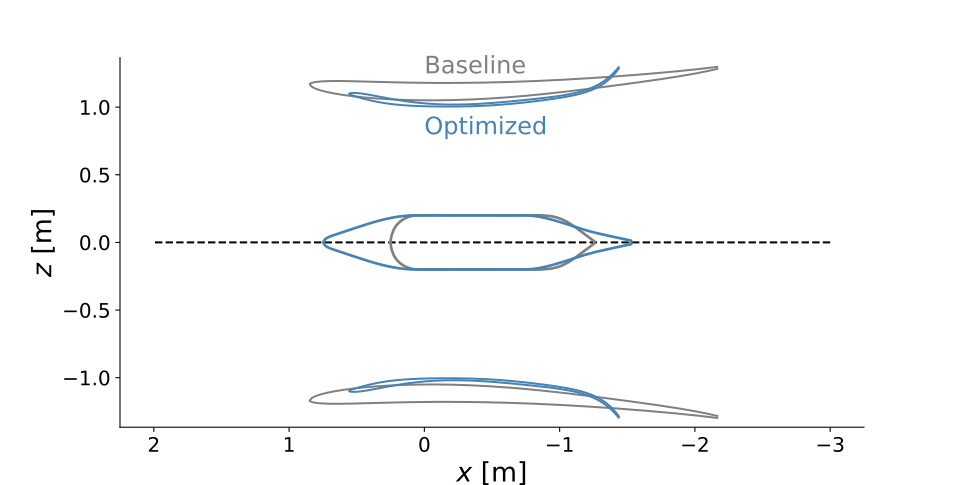}}
\caption{Comparison of the optimized duct and hub with the baseline design.}
\label{fig:results:optDuctHub}
\end{figure}

\begin{figure}[h!]
\centering
\begin{subfigure}[t]{0.58\textwidth}
    \centering
    \includegraphics[width=\textwidth]{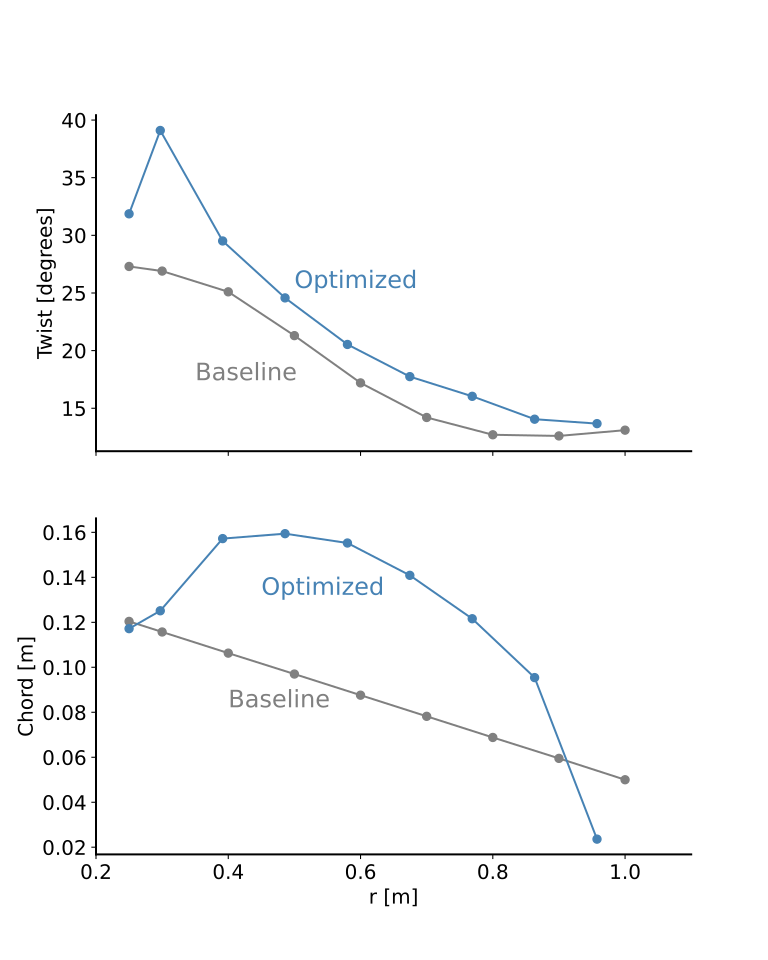}
    \caption{Twist and chord profiles}
    \label{fig:results:optBlade1}
\end{subfigure}%
\begin{subfigure}[t]{0.42\textwidth}
    \centering
    \includegraphics[width=\textwidth]{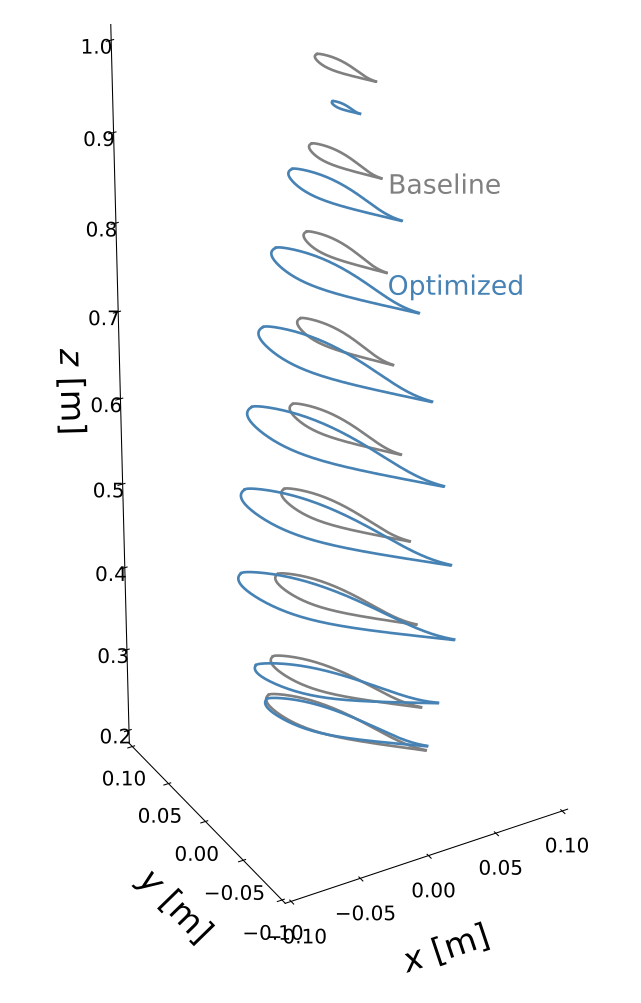}
    \caption{Spanwise sections}
    \label{fig:results:optBlade2}
\end{subfigure}
\caption{Comparison of the optimized blade with the baseline design.}
\label{fig:results:optBlade}
\end{figure}

Figure~\ref{fig:results:optDuctHub} compares the optimized geometries of the hub and duct to the baseline, and Figure~\ref{fig:results:optBlade} makes the same comparison for the blades.
The duct section evolves from a thick foil shape to a thin, highly cambered profile with a slightly rounded leading edge.
The leading edge curvature constraint of $\rho_{\kappa}=0.0015~\text{m}$ is active.
The thickness constraint of $t_{\text{duct}}=0.0014~\text{m}$ is also active away from the leading edge, with a smooth transition to thinner airfoil sections enabled by the ESP parametrization.
The optimization favors a thin-walled duct over a duct with larger thickness because for fixed outlet area $A$, a thin-walled structure allows for a more cambered profile and a larger blade radius $R_b$ (in our case, $R_b$ decreases only slightly from $1~\text{m}$ to $0.958~\text{m}$).
Physically, a higher camber and higher cone angle $\alpha$ (also observed in the optimized design) generate a stronger circulation in the duct.
This increases flow acceleration through the duct and, ultimately, a higher flow rate across the turbine blades, which increases turbine performance.  
The relation between performance and flow rate is discussed in our previous work~\cite{park2023cfd}.
The optimized design shows a tradeoff between the flow acceleration and reduced blade radius, both direct consequences of the higher camber on a fixed-outlet-area configuration.

The optimized hub features an elongated shape, with its front end protruding in front of the duct inlet.
This configuration moves the stagnation point at the hub front further upstream, allowing more streamwise distance for flow acceleration before reaching the turbine blades.
For the turbine blades, the optimized design exhibits substantially increased chords, particularly between $0.4R$ and $0.8R$, where the majority of power is extracted.
The section twist angles increase over the span, with more significant changes occurring near the hub to accommodate the bulky hub's influence.
Such a large deformation is due to the baseline blade design not considering hub effects~\cite{park2023cfd}. 

Finally, the section closest to the hub shows a small deviation from the baseline, likely to limit mesh deformations at the blade-hub connection. These deformations negatively affect the convergence in the RANS and discrete adjoint solvers.
To mitigate this issue, we added tight bounds on the variation of this section for each optimization iteration, leading to smaller changes in the twist and chord compared to nearby sections.
These constraints are gradually relaxed with each optimization restart and are ultimately not active in the final design.
Nevertheless, this approach may limit design space exploration for optimization robustness.
Future work with more robust meshing deformation algorithms could explore and leverage larger chord and twist deflection, potentially further improving the ducted turbine's performance.

\subsection{Re-evaluation}
\label{subsec:results:reeval} 

\begin{figure}[h!]
\centering
\begin{subfigure}[t]{0.49\textwidth}
    \centering
    \includegraphics[width=\textwidth]{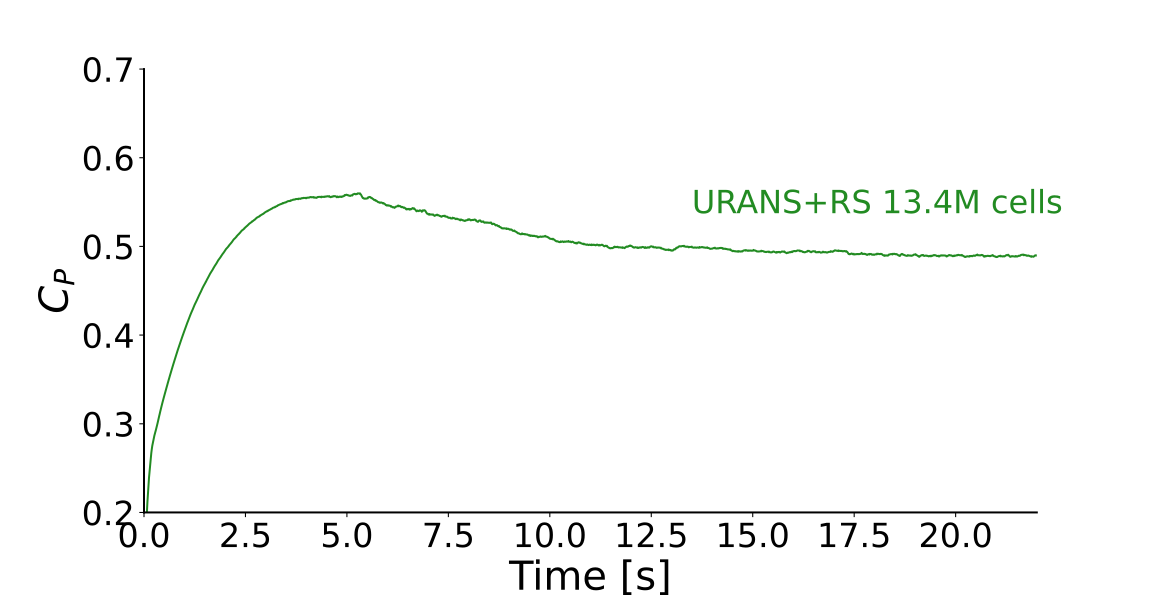}    
\end{subfigure}%
\hfill
\begin{subfigure}[t]{0.49\textwidth}
    \centering
    \includegraphics[width=\textwidth]{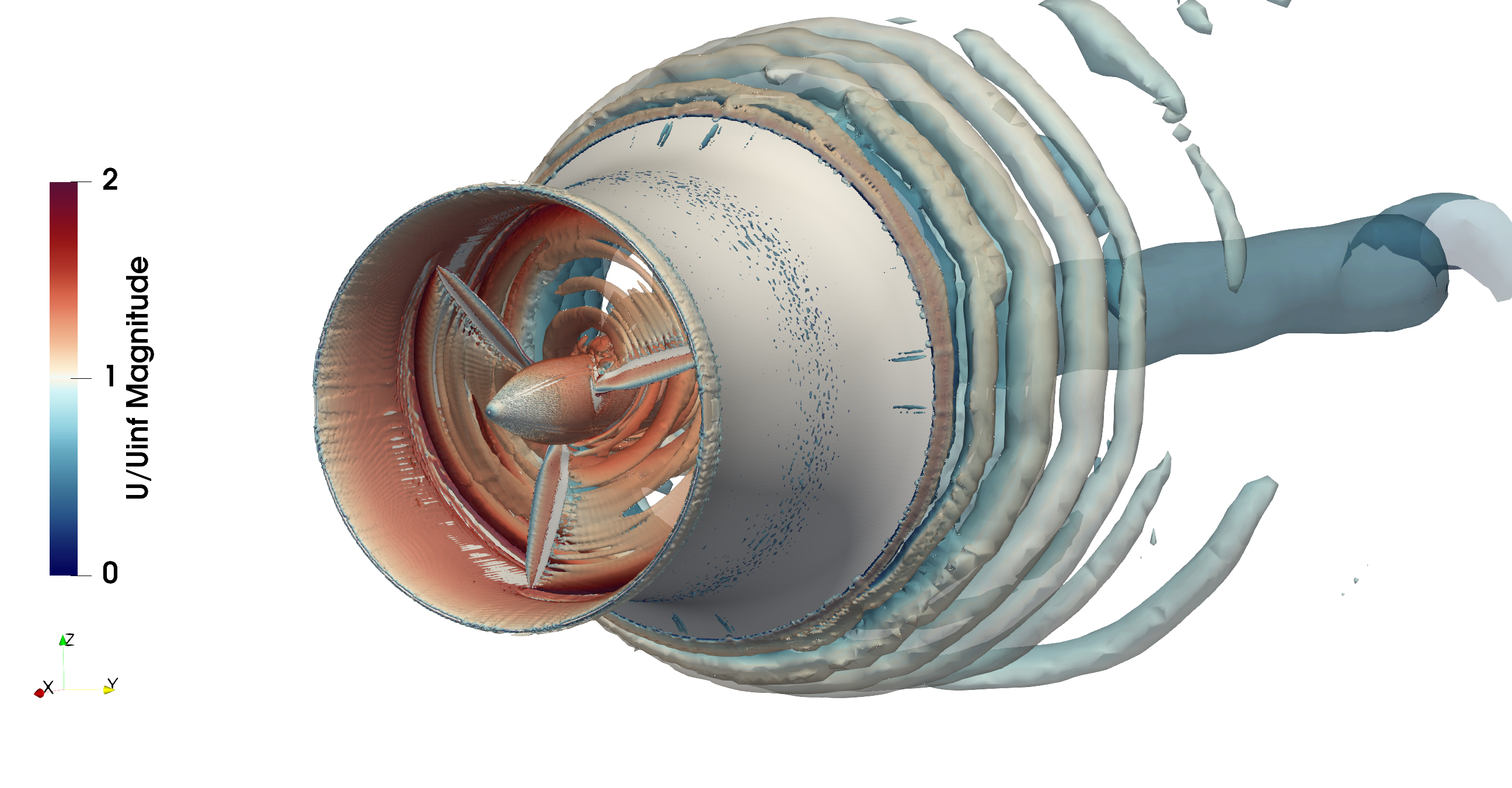}    
\end{subfigure}
\caption{Power coefficient $C_P$ as a function of time during the URANS-RS simulation using the mesh with 13.4 million cells (left) and the corresponding flow field, illustrating 3D vortex structures using Q-criterion ($Q=0.2$) at the stationary state of performance colored by normalized velocity magnitude (right).}
\label{fig:results:URANS_convergence}
\end{figure}

In this section, we re-evaluate the performance of the optimized design using a higher-fidelity URANS-RS method as discussed in Section~\ref{sec:Method}.
These URANS simulations for ducted turbines are more computationally expensive than those for freestream turbines because ducted turbines take a longer time to reach a stationary state of performance.
A typical simulation on a mesh with 13.4M cells requires at least $18~\text{s}$ of simulated physical time, equivalent to approximately $24$ turbine rotations, as shown in Figure~\ref{fig:results:URANS_convergence}. 
With an adaptive time step averaging $10^{-4}$ seconds (rotating $0.04^{\circ}$ per time step), the simulation takes approximately 10 days and 20 hours of wall time using 256 cores on the same cluster.

Table~\ref{table:results:URANS} summarizes the URANS results obtained using four meshes with different numbers of cells.
When increasing the number of cells, the value of $C_P$ converges to a range between $0.48$ and $0.49$, consistent with the RANS simulation despite exhibiting oscillatory behavior.
Considering both RANS and URANS results, a value of $C_P$ up to $50\%$ is expected for the optimized design.
This value of $C_P$ is considerably higher than the maximum that can be achieved by the freestream turbine over a range of $\lambda$.
The freestream turbine with the bulky baseline hub achieved a $C_P$ of about $45\%$ according to the RANS and URANS results shown in Figure~\ref{fig:results:bahaj}.
The optimization of the freestream turbine does not provide a significant performance improvement~\cite{park2023cfd}, but adding a duct improves the turbine performance even when the hub is present.
Moreover, the overall $C_T$ value for the entire ducted turbine is between $0.92$ and $0.96$, as shown in Table~\ref{table:results:URANS}.
This is higher than for the freestream turbine's $C_T=0.838$ (see Figure~\ref{fig:results:bahaj}).
The higher $C_T$ value indicates an increased total force on the system. However, the $C_T$ on the rotor itself is substantially reduced in the ducted turbine.
This may be advantageous when anchoring the device on-site because it lowers the force directly acting on the rotor and shifts more of the load onto the surrounding structure.

\begin{table}[h!]
\centering
\begin{adjustbox}{width=0.7\columnwidth,center}
\begin{tabular}{rrrccc}
\hline
     \textbf{Cells} & \textbf{$y^{+}$ Blade} & \textbf{$y^{+}$ Duct} & \textbf{$C_P$} & \textbf{$C_{T,blade}$} & \textbf{$C_{T,duct}$} \\
    \hline
     4.4M & 45.0 & 118.9 &0.450 &0.428 &0.474\\
     7.9M & 34.2 & 74.8 &0.485 &0.446 &0.503\\
     13.4M & 27.0 & 55.4 &0.490 &0.451 &0.508\\
     19.7M & 22.3 & 46.2 &0.480 &0.443 &0.487\\
\hline\hline
\end{tabular}
\end{adjustbox}
\caption{URANS-RS re-evaluation results of the optimized design with different mesh resolutions, including cell counts, $C_P$, $C_T$, as well as average $y^+$ values on both duct and blades. These $y^+$ values are within the acceptable range for the automatic wall treatment used in our study.}
\label{table:results:URANS}
\end{table}

The value of $C_P$ from the RANS simulations of the current design (0.501) is higher than that of the design from our previous work~\cite{park2023cfd} (0.482). However, 
the URANS results show a different trend, with $C_P=0.48\sim0.49$ and $0.54\sim0.55$ for the current and previous designs, respectively. The uncertainty in CFD simulations due to mesh and modeling errors must be taken into account when comparing these two cases. 

Besides this uncertainty, two factors in the current design could negatively affect the performance in terms of $C_P$.
The first factor is the bulky hub, which tends to reduce the flow velocity and occupy the space that could otherwise be used by the blades for power extraction. 
The second factor is the bound on the leading edge radius of curvature.
The rounded leading edge tends to improve the performance at off-design conditions (e.g., oblique inflow) with some sacrifice of the performance at the on-design condition~\cite{Madsen2019a,Mangano2021a}.
Considering the additional constraints and hub negatively affecting the peak performance, the current design's performance exhibiting considerably higher $C_P$ compared to the unducted counterpart should be applauded. This achievement showcases once more the hydrodynamic benefits of ducted turbine configuration.

\begin{figure}[h!]
\centerline{\includegraphics[width=0.75\textwidth]{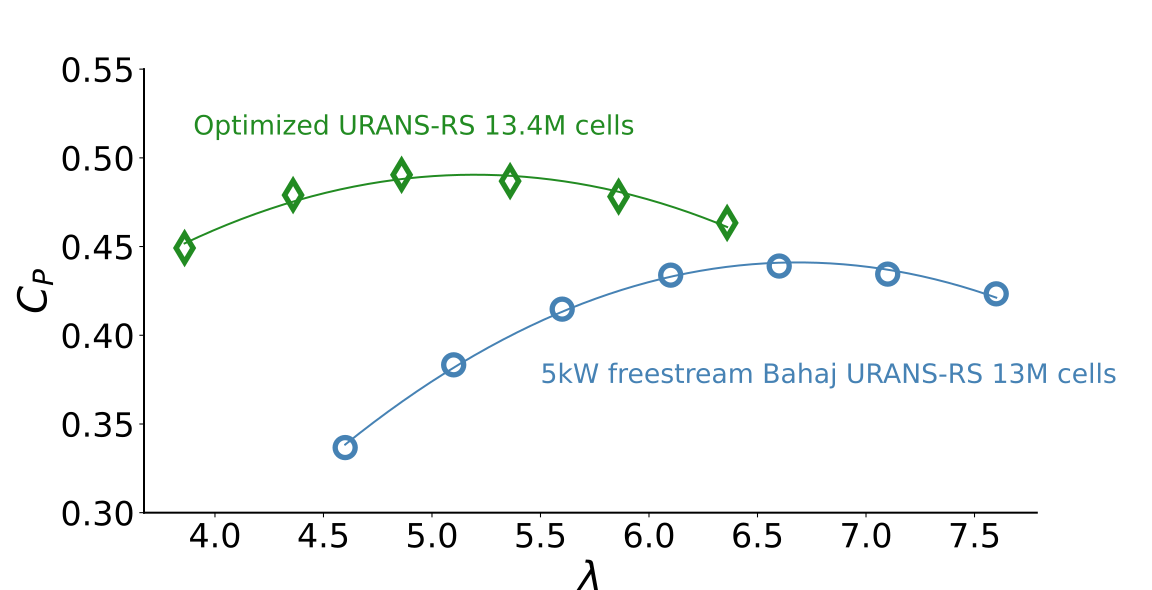}}
\caption{Comparison of power coefficient $C_P$ between the optimized ducted turbine and the $5~\text{kW}$ freestream Bahaj with the baseline hub over a range of $\lambda$, using a mesh with a similar number of cells.} 
\label{fig:results:sweep}
\end{figure}

Figure~\ref{fig:results:sweep} plots the values of $C_P$ obtained with the URANS solver using the 13.4M cell mesh for a range of $3.86\leq\lambda\leq6.36$.
The ducted turbine outperforms the $5~\text{kW}$ Bahaj turbine over the whole $\lambda$ range despite considering a fixed $\lambda=4.86$ during the optimization.
This further supports the advantage of the optimized ducted turbine in realistic situations with a time-varying tip speed ratio.

\section{Conclusions}
\label{sec:conclusion}

In this study, we conduct a gradient-based design optimization of a $5~\text{kW}$ ducted hydrokinetic turbine using CFD and the adjoint method.
We optimize a total of 37 design variables, parametrizing the duct, blade, and hub geometries.
We built this parametrization with ESP and used it to construct practical geometric constraints, including a minimum hub size to house the generator and a minimum duct thickness for manufacturing and structural considerations.

Our optimized design yields significantly improved performance relative to the baseline design at a broad range of $\lambda$, achieving a $C_P$ of up to $50\%$.
The optimized design's features include a hub protruding from the inlet of the duct, a highly cambered thin-walled duct constrained only by the thickness and leading edge curvature constraints, and blades with increased twist near the base and enlarged chords in their midspan. 
All these features have different physically beneficial effects on the turbine's performance.

This work demonstrates the effectiveness of using gradient-based optimization with practical geometry constraints in the design of high-efficiency ducted hydrokinetic turbines.
In future work, the optimization problem could be extended to include supporting structures for the hub, either as constraints or additional design variables.
Additionally, we completed an experimental study of a ducted turbine developed with our optimization framework in the University of Michigan's Aaron Friedman Marine Hydrodynamics Laboratory~\cite{naik4973480experimental}.
These efforts aim to push the boundaries of performance in ducted turbine systems and enhance their viability for renewable energy applications.

\section*{Acknowledgements}
This work was supported by the US Department of Energy under the award ``RAFT: Reconfigurable Array of High-Efficiency Ducted Turbines for Hydrokinetic Energy Harvesting'' (Award No. DE-AR0001438). The authors thank DOE ARPA-E Submarine Hydrokinetic And Riverine Kilo-megawatt Systems (SHARKS) Program led by Mario Garcia-Sanz. We also thank the entire RAFT Team, especially Onur Bilgen, Gregory Methon, and Nazim Erol from the Structural team, Yue Cao, Md Tariquzzaman and Peidong Li from the Electrical team, and Ruo-Qian (Roger) Wang and Eshwanth Asok from the Environmental team, for their valuable discussions during the work. Help from Ping He on DAFoam is also greatly appreciated. This work utilized the FASTER high-performance computing cluster at Texas A\&M University through allocation PHY220115. This computing allocation is supported by the Advanced Cyberinfrastructure Coordination Ecosystem: Services \& Support (ACCESS) program, which is funded by National Science Foundation grants \#2138259, \#2138286, \#2138307, \#2137603 and \#2138296.

\appendix
\section{Foil shape parametrization using Class-Shape Transformation (CST)}
\label{app:cst}

The class-shape transformation (CST)~\cite{kulfan2006fundamental} parametrization provides a flexible and efficient representation of complex geometries, such as foils in our case, through the multiplication of class and shape functions.
The class function $C$ defines the overall shape category or ``class" of geometry, given in generic form by:
\begin{equation}
\label{app:eq:classfunc}
    C(\psi) \equiv \psi^{N_1}(1-\psi)^{N_2},
\end{equation}
where the coordinate $\psi$ varies from 0 to 1, and the exponents $N_1$ and $N_2$ define the type of geometry. In the case of foil shape, $N_1=0.5$ and $N_2=1$ are typically used to create a round leading edge and a pointed trailing edge.

The detailed geometry is further refined by multiplying a shape function on the class function.
A family of well-behaved analytical functions is commonly chosen to generate the shape function $S$, for example, the Bernstein basis polynomials used in our case:
\begin{equation}
\label{app:eq:shapefunc}
    S(\psi) = \sum_{i=0}^{n} A_{i} \cdot \binom{n}{i} \psi^i (1 - \psi )^{n-i},
\end{equation}
where $n$ is the polynomial degree, $A_i$ is the Bernstein coefficient (or CST variables), and $\binom{n}{i}$ is the binomial coefficient.

Under the CST framework, the duct upper and lower surfaces are described as the product of the class and shape functions:
\begin{equation}
\label{app:eq:ductCST}
\begin{cases}{}
        \xi_{upper}(\psi) = \sqrt{\psi} (1 - \psi ) \cdot \sum_{i=0}^{n} A_{i,upper} \cdot \binom{n}{i} \psi^i (1 - \psi )^{n-i}+\psi\Delta\xi_{upper},\\
        \xi_{lower}(\psi) = \sqrt{\psi} (1 - \psi ) \cdot \sum_{i=0}^{n} A_{i,lower} \cdot \binom{n}{i} \psi^i (1 - \psi )^{n-i}+\psi\Delta\xi_{lower}.
\end{cases}
\end{equation}
Here, $\psi$ and $\xi$ are the foil-surface coordinates along the chord and thickness directions, respectively, normalized by the chord length, and $\Delta\xi$ specifies the trailing edge thickness, set at 0.5\% of the chord length in this work. This CST parametrization offers systematic adjustments to the upper and lower surfaces of the duct, ensuring smooth geometries throughout the design process.

Another advantage of using the CST variable is that the leading-edge radius of curvature can be directly expressed as
\begin{equation}
\label{app:eq:LEcurv}
    \tilde{\rho_{\kappa}}=\frac{1}{2}A_0^2,
\end{equation}
for both upper and lower surfaces, where $\tilde{\rho_{\kappa}}$ is the radius of curvature normalized by the chord length. This is computed by first substituting the first and second derivatives of Eq.~\ref{app:eq:ductCST} into the general equation for radius of curvature, i.e., $\tilde{\rho_{\kappa}}(\psi)=\{1+[\xi^{\prime}(\psi)]^2 \}^{3/2}/\xi''(\psi)$, and then taking the limit as $\psi\rightarrow0$. For a detailed derivation, readers can refer to the appendix of the original paper~\cite{kulfan2006fundamental}.


\section*{Data Availability}
All data are included in the paper, with a link to the Git repository for key files provided: \url{https://github.com/jbpark94/Ducted-Hydrokinetic-Turbine-Key-Files.git}.

\newpage
\bibliographystyle{elsarticle-num-names} 
\bibliography{ref.bib}






\end{document}